\documentclass[a4paper,11pt]{article}

\usepackage{jcappub} 
\usepackage[T1]{fontenc} 

\title{Gravitational radiation reaction around a static black hole surrounded by a Dehnen type dark matter halo}

\author[a]{Amjad Ashoorioon}
\author[b,c,d]{Roberto Casadio}
\author[a]{Khadije Jafarzade}
\author[e,a]{Mohammad B. Jahani Poshteh}
\author[e,f,g,h]{Orlando Luongo}

\affiliation[a]{School of Physics, The Institute for Research in Fundamental Sciences (IPM), P.O. Box 19395-5531, Tehran, Iran}
\affiliation[b]{Dipartimento di Fisica e Astronomia, Alma Mater Universit\`a di Bologna, 40126 Bologna, Italy}
\affiliation[c]{Istituto Nazionale di Fisica Nucleare, I.S.~FLaG, Sezione di Bologna, 40127 Bologna, Italy}
\affiliation[d]{Alma Mater Research Center on Applied Mathematics (AM$^2$), Via Saragozza 8, 40123 Bologna, Italy}
\affiliation[e]{Universit\`a di Camerino, Divisione di Fisica, Via Madonna delle carceri 9, 62032 Camerino, Italy}
\affiliation[f]{Istituto Nazionale di Fisica Nucleare (INFN), Sezione di Perugia, Perugia, 06123, Italy}
\affiliation[g]{INAF - Osservatorio Astronomico di Brera, Milano, Italy}
\affiliation[h]{Al-Farabi Kazakh National University, Al-Farabi av. 71, 050040 Almaty, Kazakhstan}

\emailAdd{amjad@ipm.ir}
\emailAdd{casadio@bo.infn.it}
\emailAdd{k.jafarzade@ipm.ir}
\emailAdd{mohammad.jahani@unicam.it}
\emailAdd{orlando.luongo@unicam.it}

\abstract{We consider the motion of a particle in the geometry of a Schwarzschild-like black hole embedded in a dark matter (DM) halo with Dehnen type density profile and calculate the orbital periods along with the evolution of the semi-latus rectum and eccentricity for extreme mass ratio inspirals (EMRIs). Such a system emits gravitational waves (GWs), and the particle's orbit evolves under radiation reaction. We also consider the effects of dynamical friction and accretion of DM on the orbital parameters. We find that the eccentricity and semi-latus rectum decrease faster with respect to the case in which EMRI is in empty spacetime.}

\begin{document}
\maketitle
\flushbottom
\newcommand{\cL}{\mathcal{L}}
\newcommand{\mpl}{M_\text{Pl}}
\newcommand{\wrt}{\textit{w.r.t.}}
\newcommand{\si}{\sigma}
\newcommand{\rdm}{\mathrm{d}}
\newcommand{\pa}{\partial}

\section{Introduction}

The interplay between black holes and DM has proved to be a significant field of study in theoretical astrophysics, especially in understanding the dynamics of gravitational systems \cite{Boshkayev:2020kle,Boshkayev:2021chc,Capozziello:2025ycu}. Black holes play a crucial role in the formation and evolution of galaxies. On the other hand, DM, which constitutes about $27 \%$ of the universe's total mass-energy content, adds further complication to these gravitational systems and in cosmology \cite{Dunsby:2016lkw}. In this context, the study of gravitational radiation reaction around black holes immersed in a DM halo offers deep insights into how these entities influence each other's behavior and the resulting GW emitted.

The presence of extended DM halos surrounding galaxies is strongly supported by various lines of evidence, including galaxy rotation curves, gravitational lensing analyses, and measurements of cosmic microwave background anisotropies. Among the models used to describe such halos, the Navarro–Frenk–White (NFW) profile \cite{Navarro:1995dmo}, derived from cold DM simulations, is widely employed and is characterized by its steep inner density cusp. Nevertheless, the NFW profile has limited adaptability for representing alternative halo morphologies, such as cored density profiles. On the other hand, the Dehnen profile \cite{Dehnen:1993} offers a more general double power-law form, encompassing the Jaffe and Hernquist models as specific cases and featuring an adjustable inner slope. This versatility makes it especially suitable for capturing both cuspy and cored DM configurations, which are relevant for modeling the environments of black holes \cite{Stegmann:2020,Hamil:2024dehnen}.
 It defines a one-parameter family of spherical equilibrium models in which the inner logarithmic density slope, $\gamma$, regulates the degree of central phase-space concentration. In contrast to fixed-shape prescriptions, such as the NFW profile, the Dehnen model provides a continuous interpolation between core-like ($\gamma \simeq 0$) and steeply cusped ($\gamma \simeq 2$) configurations without modifying its analytic form. This parametric freedom endows $\gamma$ with a physically interpretable role as an entropy-compression index, enabling the model to encapsulate a wide range of halo morphologies within a unified framework. Moreover, the model retains full analytic tractability in its cumulative mass, potential, and distribution function, rendering it particularly advantageous for dynamical modeling and gravitational-lensing applications while preserving a direct connection to the system's inner information content.

Recent advances in astrophysical research have increasingly highlighted the relevance of the Dehnen DM halo model. Since its introduction as a spherically symmetric density profile, the Dehnen model has gained widespread application in galactic dynamics and black hole astrophysics \cite{Dehnen:1993}. Initially employed to characterize the nuclear bulges of elliptical galaxies, it has more recently been used to construct black hole models embedded within Dehnen-profiled DM halos, revealing the impact of DM on the gravitational environment of black holes \cite{Pantig:2022dehnen}. Further studies have derived black hole solutions for specific Dehnen profiles, systematically analyzing their thermodynamic behavior and strong-field gravitational signatures \cite{Gohain:2024dark}. Investigations have also demonstrated that the gravitational lensing properties and orbital dynamics predicted by this model align well with general relativity, making it a preferred theory for explaining the peripheral environment of galaxies \cite{Ali:2025dehnen}. Collectively, these results establish the Dehnen profile as a powerful framework for probing DM–black hole interactions and the structure of galactic environments.

In addition, GWs have become of large interest in gravitational physics since its detection by LIGO/Virgo collaboration in 2015 \cite{LIGOScientific:2016aoc}. One of the main sources of GW is two body systems in which a stellar-mass compact object (SCO) rotates a supermassive black hole (SMBH). In such extreme mass ratio inspirals (EMRIs), low frequency GW would be produced which could be observed in future space-based observatories like LISA \cite{LISA:2022kgy}.

In this paper, we are interested in EMRIs immersed in a DM halo whose density distribution is given by Dehnen type profile \cite{Dehnen:1993uh}. Dehnen distribution has been used, by several authors, to model the DM profile of dwarf galaxies \cite{Stegmann:2019wyz, Inoue:2017voc, errani2018systematics}.  This dark-matter background serves as a generalization of the well-established NFW profile, exhibiting a partially phenomenological character. In the following, we provide theoretical motivations for adopting this form. In addition, the traditional belief is that these galaxies rarely host a SMBH at their centers. However, some recent observations, including that of Leo I galaxy, show that SMBHs might be present at the center of some dwarf galaxies \cite{Bustamante-Rosell:2021ldj} (see also Ref. \cite{Reines:2022ste}). This has motivated study of black holes embedded in Dehnen type DM halo \cite{Pantig:2022whj, Al-Badawi:2024asn, Gohain:2024eer, Al-Badawi:2024qpt, Al-Badawi:2025njy}. As the SCO spiral into the SMBH, it emits GWs. Further, the DM halo exerts a gravitational drag force on SCO which is known as dynamical friction. On the other hand, SCO accretes DM as it is moving inside the halo. We are interested in studying the evolution of the orbital parameter of SCO as well as its energy and angular momentum due to GW radiation reaction, dynamical friction, and the accretion. Various black hole solution surrounded by DM halo have been found in literature \cite{Cardoso:2020iji, Cardoso:2021wlq, Cardoso:2022whc, Figueiredo:2023gas} and the evolution of orbits of small compact objects inspiralling around these black holes has been investigated \cite{Dai:2021olt,Zhang:2024ugv}. These studies could help us find out about the existence and properties of DM.

The paper is structured as follows. In Sec. \ref{sezione2}, we describe the main features of dark matter density profile, computing the corresponding mass associated with dark matter. In Sec. \ref{sezione3}, the bound orbits of the underlying metric is explored. In Sec. \ref{sezione4}, the gravitational waveforms are studied, and finally in Sect. \ref{sezione5}, we sketch our conclusions and perspectives. 

\section{Dark matter density profile and metric function}\label{sezione2}

First, we briefly review  the mass distribution of a Dehnen type DM halo. In a spherically symmetric spacetime, the mass distribution is determined by the density profile $\rho$ through
\begin{equation}
	M_{D}=4\pi \int\limits_{0}^{r}\rho \left( r^{\prime }\right) r^{\prime
		2}dr^{\prime }.
	\label{eq:Md}
\end{equation}
The density profile of the Dehnen DM halo is a special case of a double power-law profile given by 
\begin{equation}
	\rho = \rho_s \left(\frac{r}{r_s}\right)^{-\gamma } \left[\left(\frac{r}{r_s}\right)^{\alpha }+1\right]^{\frac{\gamma -\beta }{\alpha }},
	\label{dens1}
\end{equation}
where $\rho_s$ and $r_s$ denote the central halo density and the halo core radius respectively, whereas $\gamma$ determines the specific variant of the profile. The values of $\gamma$ lies within
$[0, 3]$, e.g.~$\gamma = 3/2$ is used to fit the surface brightness profiles of elliptical galaxies which closely resembles the de~Vaucouleurs $r^{1/4}$ profile \cite{Shakeshaft:1974iau}.

 In general, the Dehnen model offers an improved representation of the dark-matter density profiles observed in $N$-body simulations. Owing to its analytic flexibility, the Dehnen family can accommodate physically motivated variations in the inner structure of halos. In particular, as demonstrated in Ref.~\cite{Gondolo:1999ef}, the adiabatic growth of a supermassive black hole (SMBH) at the halo center transforms an initial power-law cusp, $\rho_i(r) \propto r^{-\gamma}$ with $0 < \gamma < 2$, into a final density spike, $\rho_f(r) \propto r^{-\gamma_{\mathrm{sp}}}$, where $2.25 < \gamma_{\mathrm{sp}} < 2.5$. The adopted Dehnen profile with $\gamma = 2.5$ thus reproduces the theoretically expected inner asymptotic behavior of such spiked halos within a fully analytic framework. By contrast, the standard NFW model--with its fixed inner slope $\gamma = 1$--is incapable of capturing this physically motivated steep cusp. The Dehnen formulation, therefore, provides a more realistic and analytically tractable description of black hole-halo systems.

In this work, we conventionally consider the Dehnen-$\left( \alpha,\beta ,\gamma \right) =\left( 1,4,5/2\right)$ DM halo, to guarantee agreement with observations. 

Accordingly, Eq.~(\ref{dens1})
becomes 
\begin{equation}
	\rho =\frac{\rho _{s}}{\left( \frac{r}{r_{s}}\right) ^{5/2}\left( \frac{r}{r_{s}}+1\right) ^{3/2}},
	\label{dens2}
\end{equation}
and, so, plugging Eq.~(\ref{dens2}) in Eq.~\eqref{eq:Md}, one obtains 
\begin{equation}
	M_{D}=4\pi \int\limits_{0}^{r}\frac{\rho _{s}r^{\prime
			2}}{\left( \frac{r^{\prime}}{r_{s}}\right) ^{5/2}\left( \frac{r^{\prime}}{r_{s}}+1\right) ^{3/2}} dr^{\prime }=\frac{8\pi \rho _{s}r_{s}^{3}}{\sqrt{1+\frac{r_s}{r}}}.
\end{equation}
The static black hole metric in the Dehnen-$(1,4,5/2)$ DM halo is therefore given by~\footnote{The authors of
	Ref.~\cite{Al-Badawi:2024asn}  model DM using an anisotropic fluid stress-energy tensor.
	Similarly, an effective anisotropic fluid has been proposed in Refs.~\cite{Cadoni:2017evg,Cadoni:2018dnd,Giusti:2021shf}
	to unify dark energy and DM. The procedure to model, with an external energy-momentum tensor, induced by DM, a background metric appears a viable approximation to characterize with \emph{one metric only}, two different spacetimes, i.e., an interior and an exterior configuration. Clearly, this limits the overall description, since it exhibits the presence of pressures violating the the cosmological cold DM hypothesis. However, it is possible to handle this technique at first approximation, albeit a more precise and complete scheme would require the Israel-Darmois junction conditions \cite{Chu:2021uec} or alternatives, see e.g. \cite{Luongo:2023aib,Luongo:2014qoa}. }~\cite{Al-Badawi:2024asn}
\begin{equation} 
	ds^{2}=-f\left( r\right) dt^{2}+\frac{dr^{2}}{f\left( r\right) }+r^{2}\left(
	d\theta ^{2}+\sin ^{2}\theta d\phi ^{2}\right) ,\label{m1}
\end{equation} 
where
\begin{equation}
	f\left( r\right) =1-\frac{2M}{r}-32\pi \rho _{s}r_{s}^{3}\sqrt{\frac{r+r_s}{r_s^2\,r}}\, .  \label{laps1}
\end{equation}
This choice, though restrictive compared to the more general ansatz $g_{tt} = -e^{\Phi(r)} f(r)$, which can describe spacetimes with anisotropic matter distributions, is consistent with the weak-field origin of the DM halo model adopted in Ref.~\cite{Al-Badawi:2024asn} and ensures compliance with standard energy conditions. Note that this geometry differs from the vacuum Schwarzschild black hole and the inclusion of DM has changed the horizon radius  (see e.g. Ref. \cite{Xu:2018wow}).

\section{Bound orbits of the corresponding black hole}\label{sezione3}

In this section, we study the bound orbits
of a Schwarzschild-like black hole embedded in a Dehnen type DM halo.

\subsection{The geodesic equations}

Timelike geodesics in a spherical symmetry spacetime
are described by the equations
\begin{eqnarray}
	&&\frac{dt}{d\tau}=\frac{E}{f(r)},\\
	&&\frac{d \phi}{d \tau}= \frac{L}{r^{2}},\\
	&&\left( \frac{dr}{d\tau}\right) ^{2}+V(r;L)=E^{2}, \label{eq:dr}
\end{eqnarray}
where $ \tau $ is the particle's proper time\footnote{ Since the analysis is restricted to the weak-field region where $f(r)\approx1$, the difference between $t$ and proper time $\tau$ is negligible. This is a standard simplification for large-radius orbits in similar studies.}; $ E $ and $ L $ are constants of motion, respectively, the orbital
energy and angular momentum; the effective potential for radial motion is given by
\begin{equation}
	V(r;L)=f(r)\left( 1+\frac{L}{r^{2}}\right).
\end{equation}
Note that due to the spherically symmetric property of the black hole, we can assume that the motion takes place on the equatorial plane 
$ \theta =\pi/2 $ without loss of generality.

\subsection{Orbital parameters: $p$ and $e$}

\begin{figure}[htp]
	\centering
	\includegraphics[width=0.45\textwidth]{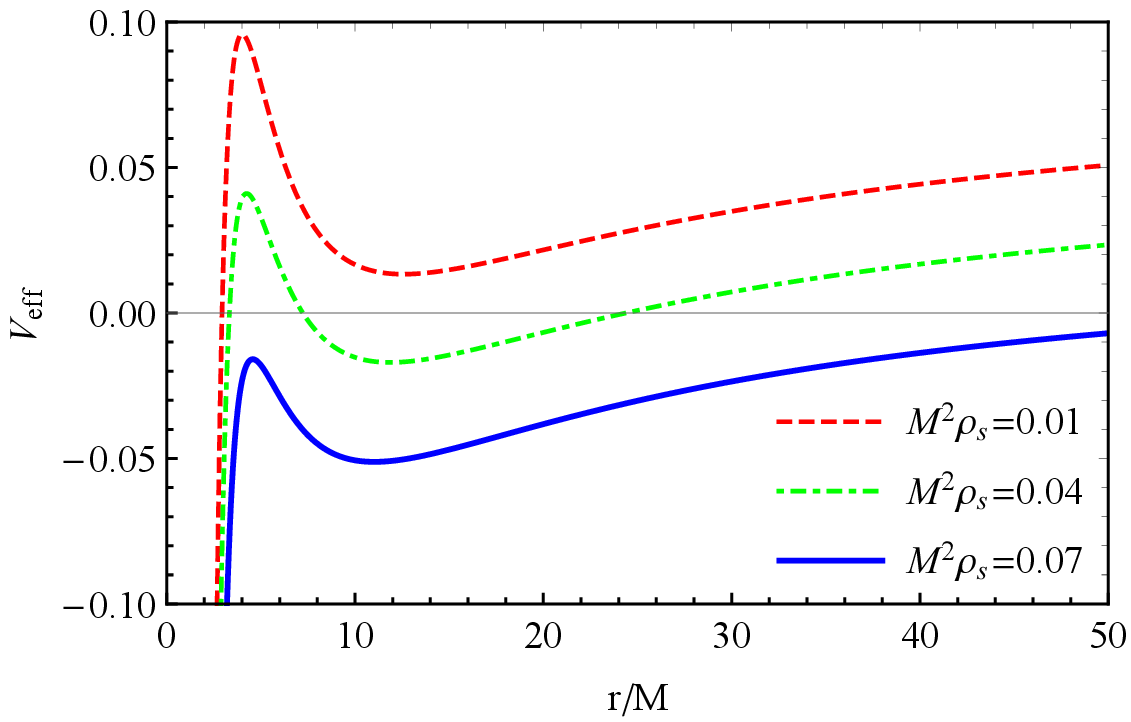}
	\includegraphics[width=0.45\textwidth]{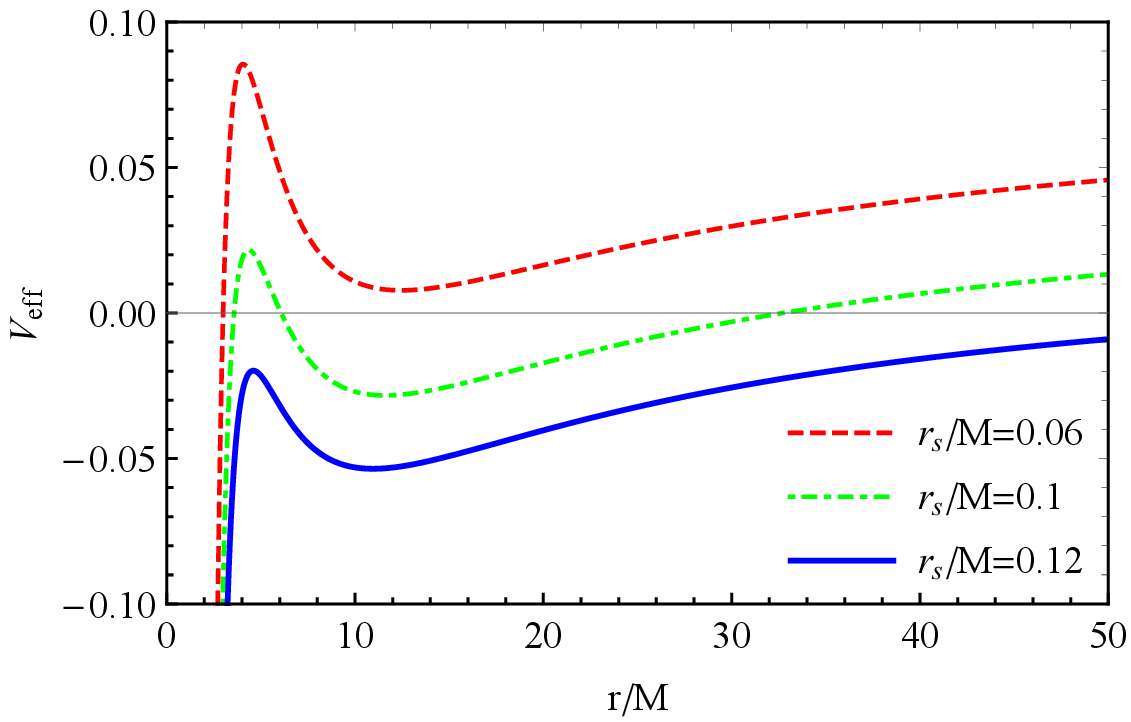}
	\includegraphics[width=0.45\textwidth]{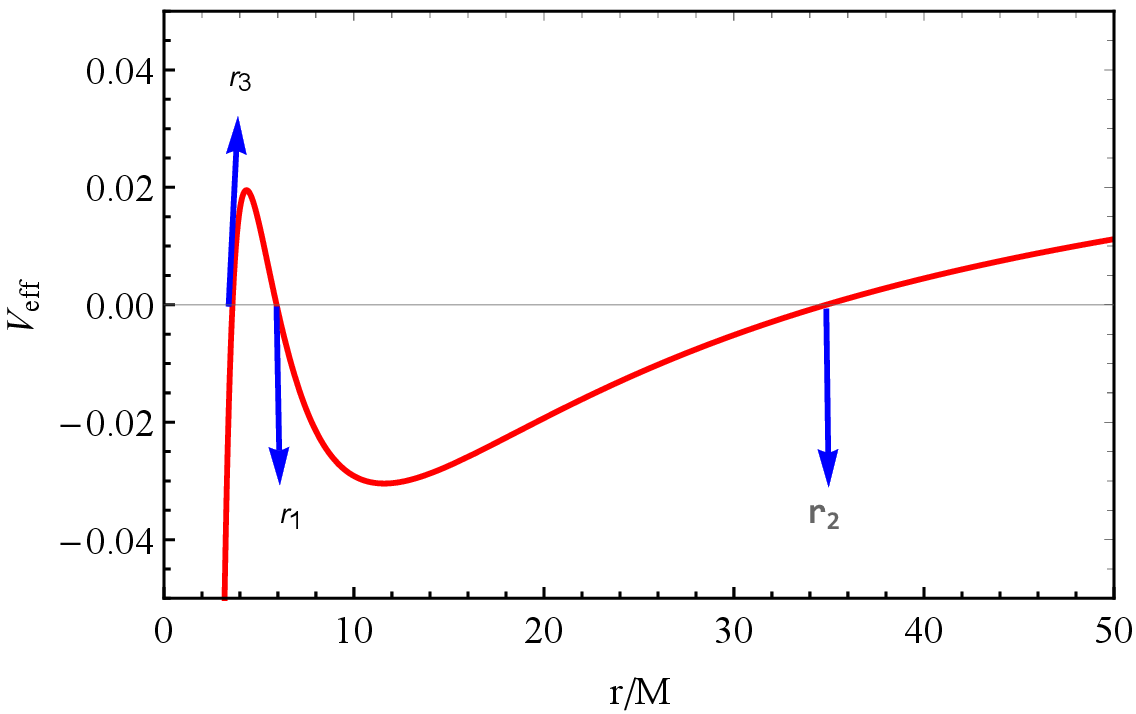}
	\caption{The effective potential for radial motion for $ r_{s}=0.1$ and different values of $ \rho_{s} $ (top panel); for 
		$ \rho_{s}=0.05 $ and different values of $ r_{s} $ (middle panel); for $ \rho_{s}=0.05 $ and $ r_{s}=0.1$ (bottom panel).}
	\label{Fig1}
\end{figure}

For bound orbits, the energy and angular momentum should be limited as $0 < E < 1$ and $ L\geq L_{\rm bound} $, where $ L_{\rm bound} $ can be obtained by determining the inflection point of the effective potential from
\begin{equation}
	\frac{\partial V}{\partial r}= \frac{\partial^{2} V}{\partial r^{2}}=0.
\end{equation}
For our case, we have
\begin{equation}
	L_{\rm bound}=\frac{2\sqrt{3}\left( M+ 8\pi \rho_{s} r_{s}^{3}\right) }{1-32 \rho_{s} r_{s}^{2}}.
\end{equation}
Clearly, for $ \rho_{s}\rightarrow 0 $ and $ r_{s}\rightarrow 0 $, one recovers $L_{\rm bound}$ of the Schwarzschild case. 
When the above constraints for bound orbits are satisfied, the cubic equation 
\begin{equation}
V(r,L)-E^{2}=0
\label{eq:cub}
\end{equation}
generally admits three distinct roots, which are designated by $r_{3} \leq r_{2} \leq r_{1}  $ as shown in Fig.~\ref{Fig1},
confirming that the existence of three roots depends on the values of DM parameters $r_{s}$ and $\rho_{s}$.


The roots of Eq.~\eqref{eq:cub} can be obtained by defining~\cite{Hughes:2024tja}
\begin{equation}
	R_3(r) = r^3 + \mathcal{A}_1r^2 + \mathcal{A}_2r + \mathcal{A}_3\;,
\end{equation}
where
\begin{eqnarray}
	\mathcal{A}_1 &=& -\frac{2M+16\pi \rho_{s} r_{s}^{3}}{(1-E^2 - 32\pi \rho_{s} r_{s}^{2})}\;,
	\\
	\mathcal{A}_2 &=& \frac{L^2(1-32\pi \rho_{s} r_{s}^{2}) }{(1-E^2 - 32\pi \rho_{s} r_{s}^{2})}\;,
	\\
	\mathcal{A}_3 &=& -\frac{2L^2(M+8\pi \rho_{s} r_{s}^{3})}{(1-E^2 - 32\pi \rho_{s} r_{s}^{2})}\;.
\end{eqnarray}
Afterwards, define
\begin{eqnarray}
	\mathcal{Q} &=& \frac{1}{9}\left(\mathcal{A}_1^2 - 3\mathcal{A}_2\right)\;,\;
	\label{eq:Q}\\
	\mathcal{R} &=& \frac{1}{54}\left(2\mathcal{A}_1^3 - 9\mathcal{A}_1\mathcal{A}_2 + 27\mathcal{A}_3\right),
	\label{eq:R}
\end{eqnarray}
and
\begin{equation}
	\vartheta = \arccos\left(\mathcal{R}/\sqrt{\mathcal{Q}^3}\right)\;.
	\label{eq:theta}
\end{equation}
Then, we have
\begin{eqnarray}
	r_{\rm 1} &=& -2\sqrt{\mathcal{Q}}\cos\left(\frac{\vartheta + 2\pi}{3}\right) - \frac{\mathcal{A}_1}{3}\;,
	\label{eq:r_a_eq}\\
	r_{\rm 2} &=& -2\sqrt{\mathcal{Q}}\cos\left(\frac{\vartheta - 2\pi}{3}\right) - \frac{\mathcal{A}_1}{3}\;,
	\label{eq:r_p_eq}\\
	r_3 &=& -2\sqrt{\mathcal{Q}}\cos\left(\frac{\vartheta}{3}\right) - \frac{\mathcal{A}_1}{3}\;.
	\label{eq:r_3_eq}
\end{eqnarray}

For a SCO inspiraling into a SMBH surrounded by a DM halo, the geodesic motion of the SCO can be parameterized by the semimajor axis $ a $ and eccentricity $e$ through \cite{Barack:2022pde}
\begin{equation}
	\label{rpe}
	r=\frac{a(1-e^{2})}{1+e \cos\chi},
\end{equation}
where $\chi$ is the orbital parameter that varies monotonically from $\chi=2k\pi$ at the periastron $r_p=p/(1+e)$, to $\chi=(2k+1)\pi$ at the apastron $r_a=p/(1-e)$. Here, the semi-latus rectum $p=a(1-e^{2})$ has dimension of length and $e$ is dimensionless and satisfies $ 0\leq e<1 $.

For eccentric orbits, it is useful to relabel the outer two roots, which define the range of the smaller body’s motion: the outermost root named as the apoapsis, and the next root as the periapsis: $r_{1}=r_{a}  $, $r_{2}=r_{p}  $. From Eqs. (\ref{eq:r_a_eq}) - (\ref{eq:r_3_eq}), the third root  can be written  in terms of $p$ and $e$. Stable eccentric orbits occur when the inequality $ r_{3}<r_{2}<r_{1} $ is satisfied, while stable circular orbits occur when  $ r_{1}=r_{2}>r_{3} $.
The relations between $\left\lbrace p,e\right\rbrace $ and $\left\lbrace E,L\right\rbrace $ can be obtained as
\begin{eqnarray}
	\label{epsilon}
	\epsilon &=& E/\mu = \sqrt{\frac{f(r_{a})f(r_{p})\left( r_{a}^{2}-r_{p}^{2}\right) }{r_{a}^{2}f(r_{p})-r_{p}^{2}f(r_{a})}},
	\\
	\label{ell}
	\ell &=& L/\mu =\sqrt{\frac{r_{a}^{2}r_{p}^{2}\left( f(r_{a})-f(r_{p})\right) }{r_{a}^{2}f(r_{p})-r_{p}^{2}f(r_{a})}}.
\end{eqnarray}

Combining Eq. \eqref{eq:dr} and Eqs. \eqref{rpe}-\eqref{ell} yields
\begin{equation}
	\begin{split}
		\label{orbital3}
		\frac{dt}{d\chi}&=\frac{\epsilon}{f(r)}\frac{dr}{d\chi}\left\{f(r)\left(-1+\frac{\epsilon^2}{f(r)}-\frac{\ell^2}{r^2}\right)\right\}^{-1/2},\\
	\end{split}
\end{equation}
\begin{equation}
	\begin{split}
		\label{orbital2}
		\frac{d\phi}{d\chi}&=\frac{\ell}{r^{2}}\frac{dr}{d\chi}\left\{f(r)\left(-1+\frac{\epsilon^2}{f(r)}-\frac{\ell^2}{r^2}\right)\right\}^{-1/2}.\\
	\end{split}
\end{equation}
The orbital period $P$ and the orbital precession $\Delta\phi$ over one period are
\begin{eqnarray}
	P&=&\int_{0}^{2\pi}\frac{dt}{d\chi}d\chi,
	\label{period}
	\\
	\Delta\phi &=&\int_{0}^{2\pi}\frac{d\phi}{d\chi}d\chi-2\pi.
	\label{phi}
\end{eqnarray}


\begin{figure*}[htp]
	\centering
	\includegraphics[width=0.45\textwidth]{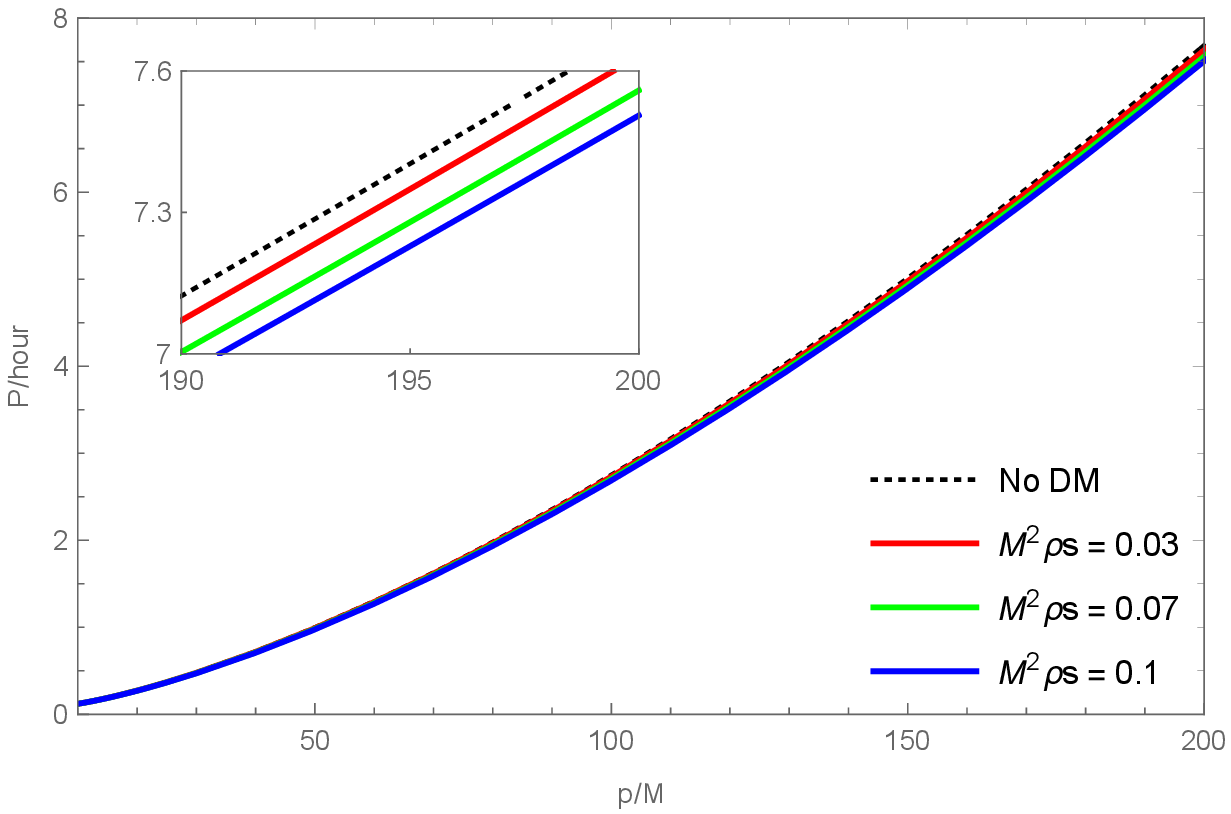}
	\includegraphics[width=0.45\textwidth]{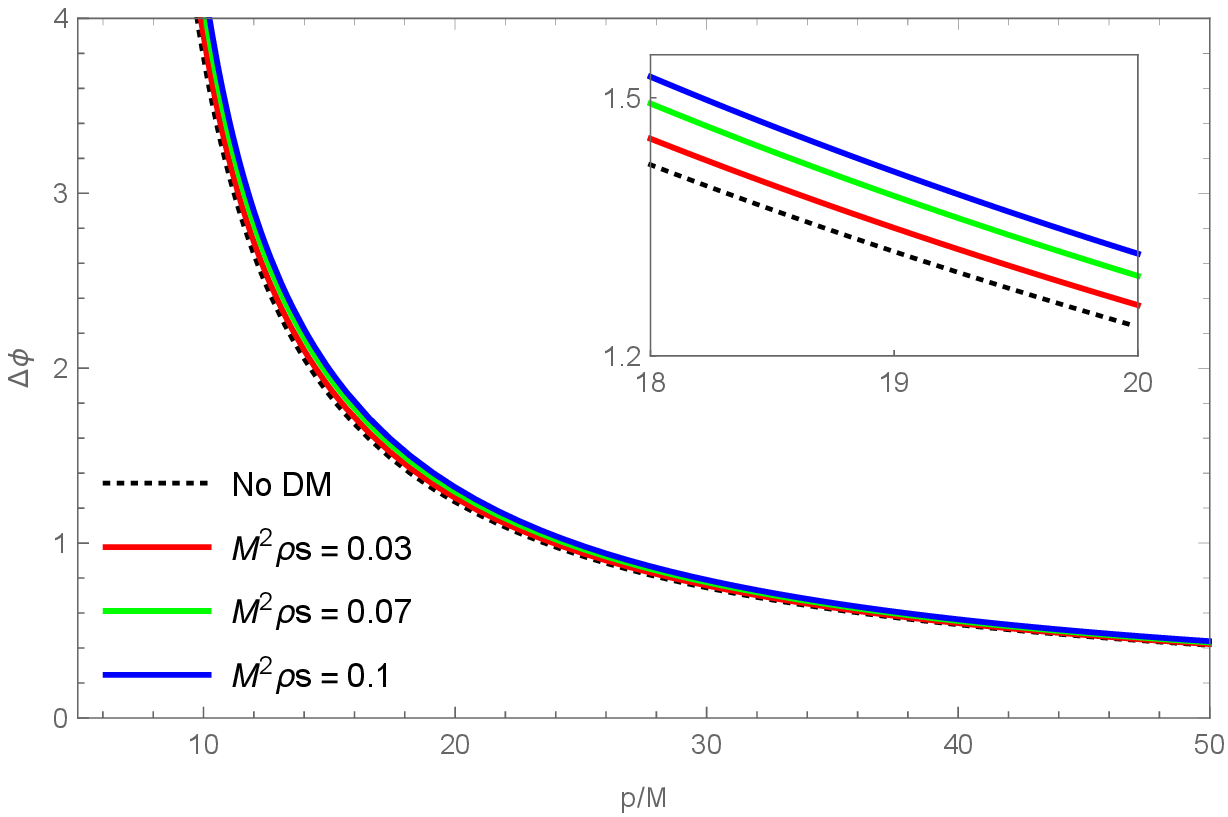}
	\includegraphics[width=0.45\textwidth]{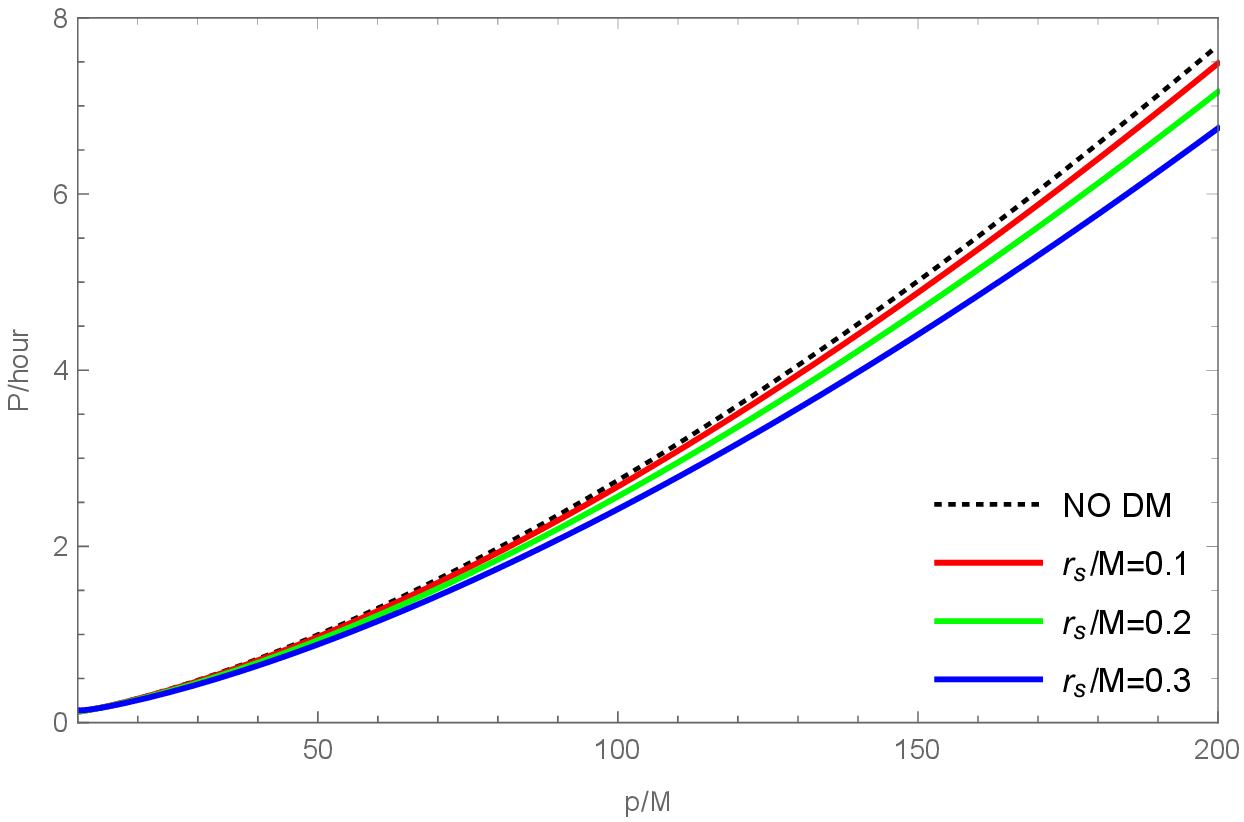}
	\includegraphics[width=0.45\textwidth]{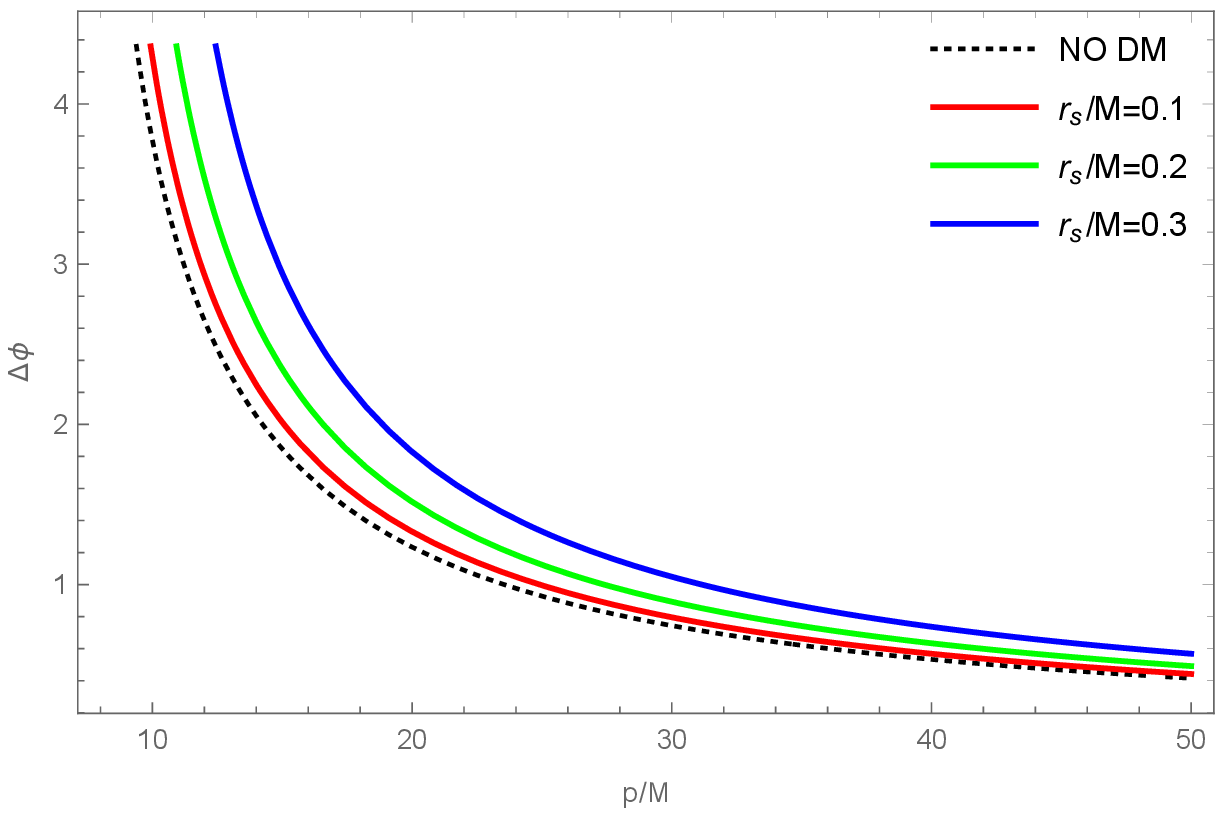}
	\caption{The results of the orbital periods and precessions for EMRIs in a Dehnen type DM halo with different values of $ \rho_{s} $ and $ r_{s}$.}
	\label{Fig2}
\end{figure*}
The influence of the halo density $\rho_s$ and scale radius $r_s$ on the orbital dynamics of an EMRI embedded in a Dehnen type DM halo is shown in Fig.~\ref{Fig2}. The presence of DM modifies both the orbital period and pericenter precession compared to the vacuum case.  As can be seen from the top panels of Fig.~\ref{Fig2}, increasing $\rho_s$ slightly decreases the orbital period and increases the pericenter precession $\Delta\phi$. The bottom panels show the dependence on the scale radius $r_s$, which controls the spatial distribution of the halo (with larger $r_s$ corresponding to a more concentrated profile), revealing that this parameter has an effect similar to that of $\rho_s$. But its effect on the orbital period is more significant at large values of $p/M$, whereas its influence on the precession is most pronounced at small $p/M$. Although these deviations are small for the parameter ranges considered, they accumulate over time and can be significant for accurate EMRI modeling in DM-rich environments.

EMRIs occurring within DM halos emit GWs and are affected by their ambient environment. The SCO's motion is shaped by several forces, including GW backreaction, dynamical friction, and material accretion from the halo.
If the SCO is a small black hole, it will draw in surrounding DM as it travels through the halo, a process that can be modeled using the Bondi-Hoyle accretion formula \cite{Macedo:2013a}
\begin{equation}
	\label{dmdt}
	\dot{\mu} = \frac{4 \pi \mu^2 \rho_\text{DM}}{(v^2+c_\text{s}^2)^{3/2}},
\end{equation}
in which $c_\text{s}=\sqrt{\delta P_\text{t}/\delta \rho_\text{DM}}$ represents the sound speed of the medium, $v$ is the velocity of the small black hole, and dots indicate time derivatives. Since accretion does not transfer orbital angular momentum, the shape of the orbit remains constant, specifically, $\left({dL}/{dt}\right)_\text{acc}=0$ and $de/dt=0$. The variation in orbital energy due to accretion can be expressed as
\begin{equation}
	\label{dEdtacc}
	\left(\frac{dE}{dt}\right)_\text{acc} =\frac{\dot{\mu}}{\mu} E+\mu\dot{\epsilon}= \frac{\dot{\mu}}{\mu} \left(E-{\ell}\frac{dp}{d\ell}\frac{dE}{dp} \right).
\end{equation}

As the small black hole moves through a DM environment, it experiences gravitational interactions with DM particles within the halo. These interactions produce dynamical friction, which contributes to the loss of both energy and angular momentum. The Chandrasekhar formula for dynamical friction is expressed as \cite{Chandrasekhar:1943}:
\begin{equation}
	\label{fDF}
	{\bf f}_{\text{DF}}=-\frac{4\pi \mu^2 \rho_\text{DM} \ln\Lambda}{v^3}{\bf v},
\end{equation}
where the Coulomb logarithm $\ln\Lambda=3$ \cite{Eda:91}. The mean rates of energy and momentum loss resulting from dynamical friction can be described as follows:
\begin{eqnarray}
	\label{dEdtdf}
	\left(\frac{dE}{dt}\right)_{\text{DF}}&=&{\bf f}_{\text{DF}}\cdot{\bf v},\\
	\label{dLdtdf}
	\left(\frac{d{\bf L}}{dt}\right)_{\text{DF}}&=&{\bf r}\times {\bf f}_{\text{DF}}.
\end{eqnarray}
The averaged rates of energy and angular momentum loss due to GW emission are expressed as \cite{Peters:131}
\begin{eqnarray}
	\label{eq:energy gw}
	\left(\frac{dE}{dt}\right)_{\text{GW}} &= &-\frac {32}{5} \frac{\mu^2}{M^2} \left( \frac{M}{p} \right)^5 f_1(e), \\
	\label{eq:angular gw}
	\left(\frac{dL}{dt}\right)_{\text{GW}} &= &-\frac {32}{5} \frac{\mu^2}{M^2} \left( \frac{M}{p} \right)^{7/2}  f_2(e),
\end{eqnarray}
where $f_1(e)$ and $f_2(e)$ are given by the following expressions
\begin{eqnarray}
	\label{eq:f1}
	f_1(e) &=&  (1-e^2)^{3/2} \left( 1 + \frac{73}{24}e^2 + \frac{37}{96}e^4\right)  , \\
	\label{eq:f2}
	f_2(e) &=&  (1-e^2)^{3/2} \left(1 + \frac{7}{8}e^2\right) .
\end{eqnarray}

The variations in orbital energy and angular momentum due to the combined influence of GW emission, accretion, and dynamical friction can be represented as:
\begin{eqnarray}
	\label{dEdtorb}
	\left(\frac{dE}{dt}\right)_{\text{orb}}&=&\left<\frac{dE}{dt}\right>_{\text{GW}}+\left<\frac{dE}{dt}\right>_{\text{acc}}+\left<\frac{dE}{dt}\right>_{\text{DF}},\\
	\label{dLdtorb}
	\left(\frac{dL}{dt}\right)_{\text{orb}}&=&\left<\frac{dL}{dt}\right>_{\text{GW}}+\left<\frac{dL}{dt}\right>_{\text{DF}}.   
\end{eqnarray}
the angle brackets denote averaging over several gravitational wavelengths, and
\begin{equation}
	\left\langle \frac{dE}{dt} \right\rangle = \frac{1}{P} \int_{0}^{P} \frac{dE}{dt}\, dt 
	= \frac{1}{P} \int_{0}^{2\pi} \frac{dE}{dt} \frac{dt}{d\phi} \, d\phi,
\end{equation}

\begin{equation}
	\left\langle \frac{dL}{dt} \right\rangle = \frac{1}{P} \int_{0}^{P} \frac{dL}{dt}\, dt
	= \frac{1}{P} \int_{0}^{2\pi} \frac{dL}{dt} \frac{dt}{d\phi} \, d\phi.
\end{equation}

\begin{figure*}[htp]
	\centering
	\includegraphics[width=0.45\textwidth]{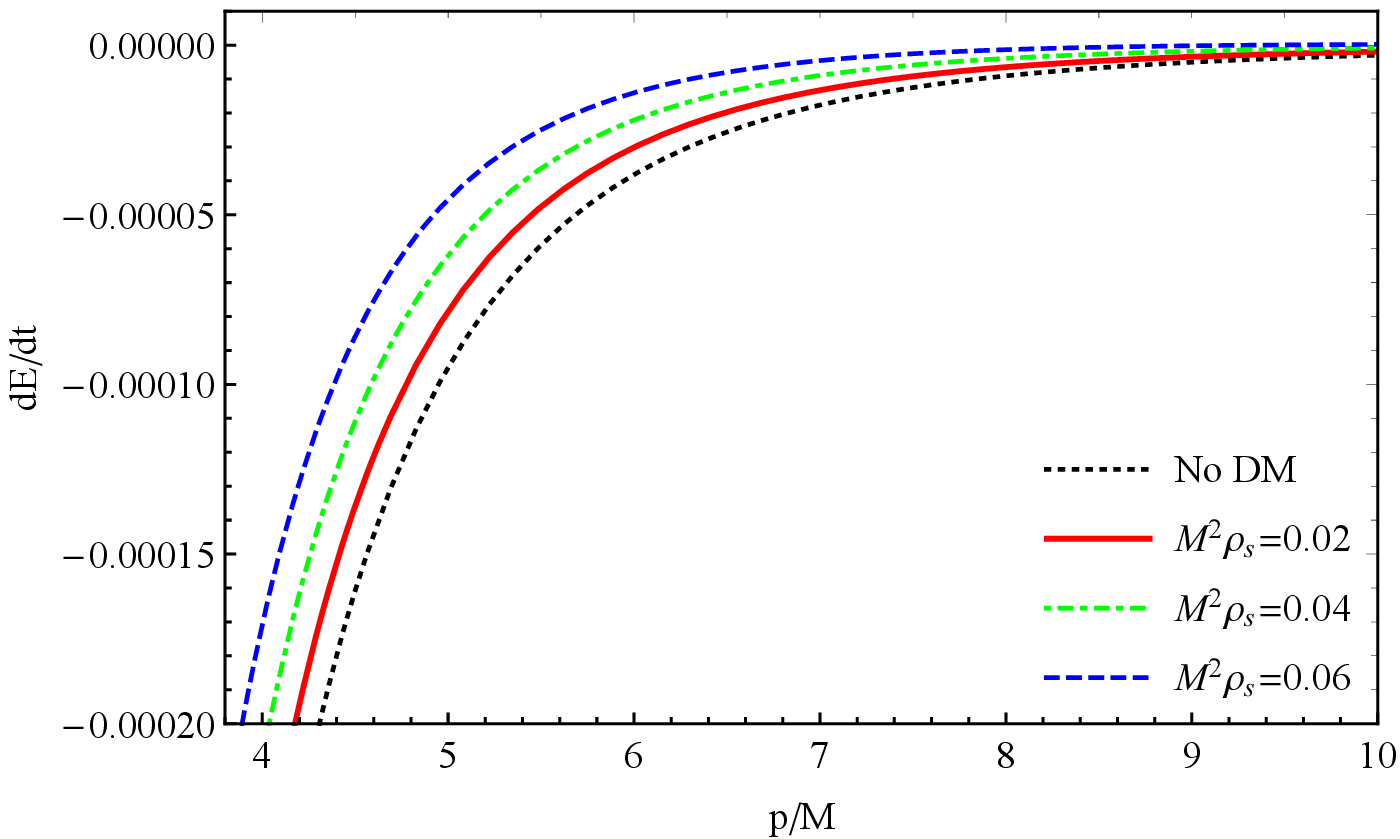}
	\includegraphics[width=0.45\textwidth]{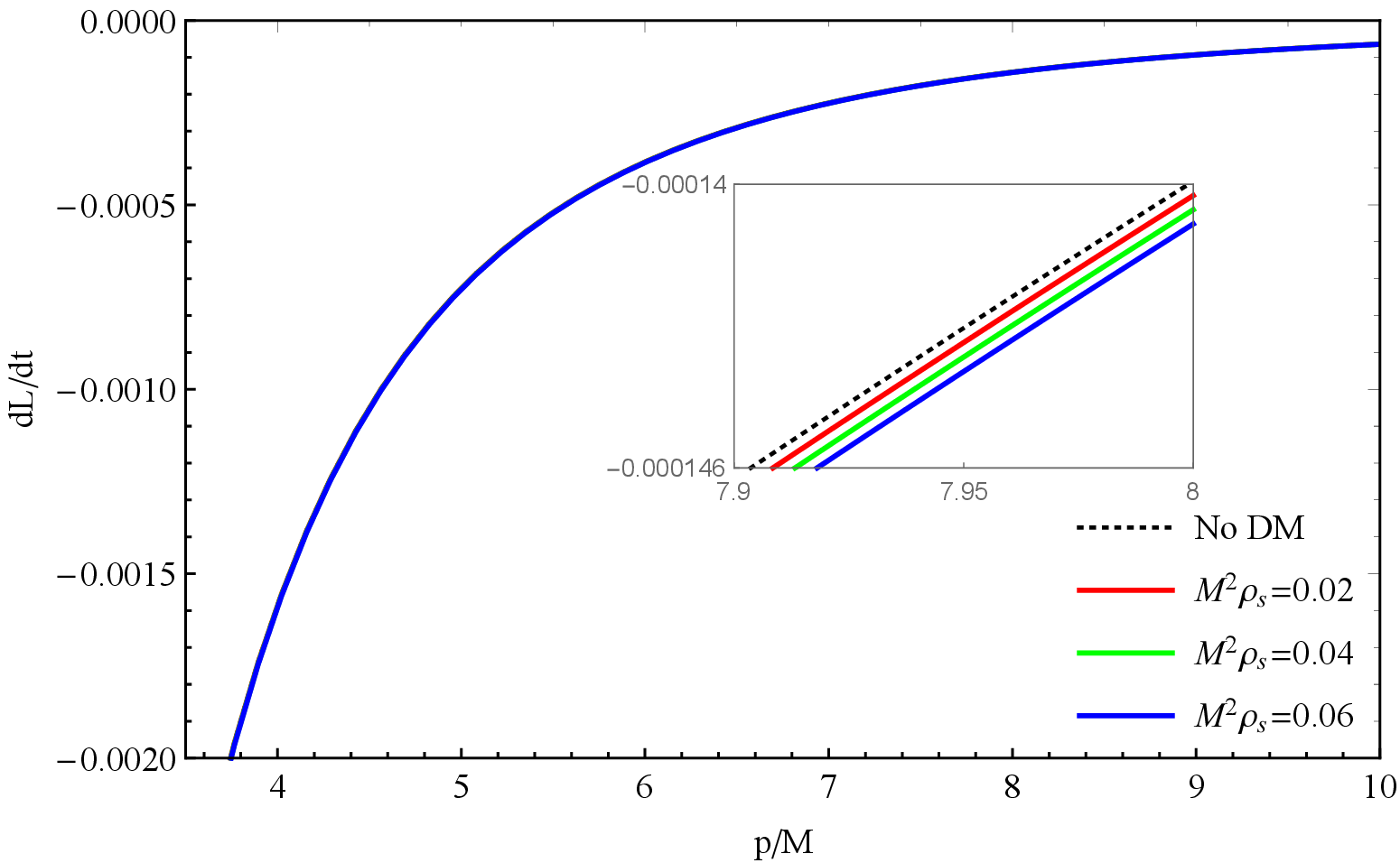}
	\includegraphics[width=0.45\textwidth]{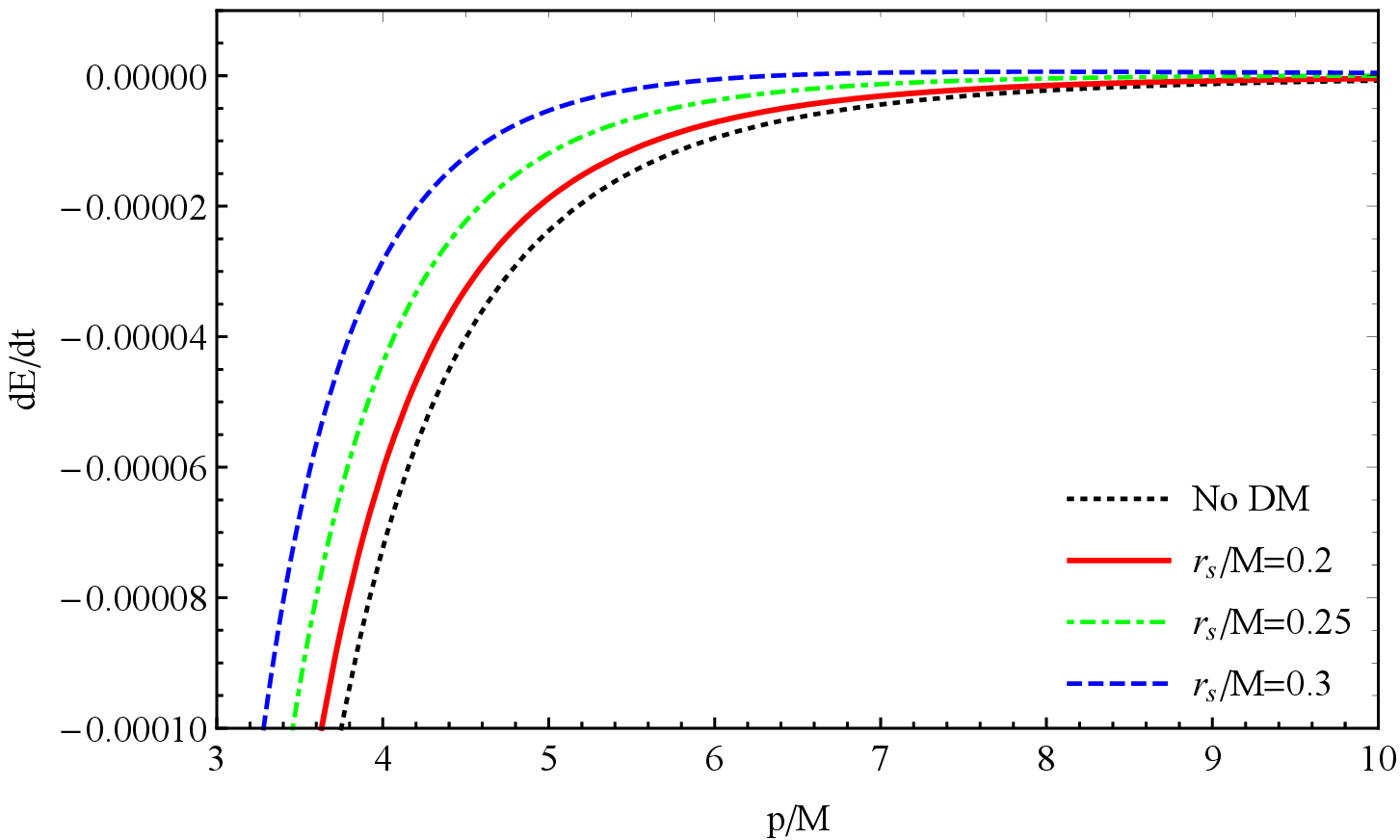}
	\includegraphics[width=0.45\textwidth]{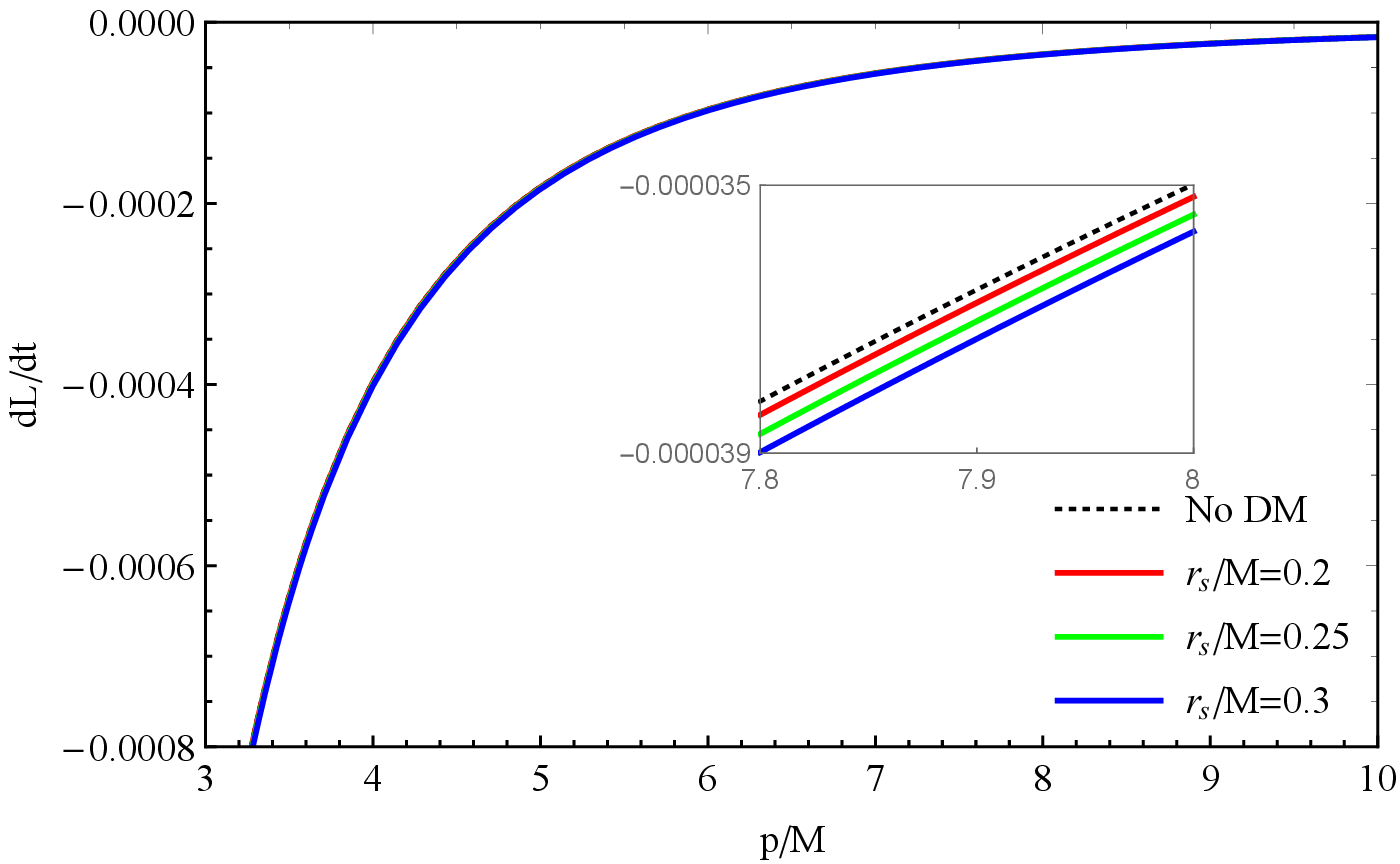}
	\caption{The energy and angular momentum fluxes for EMRIs in galaxies with a Dehnen type DM halo with different values of $ \rho_{s} $ and $ r_{s}$.}
	\label{Fig3}
\end{figure*}

Fig.~\ref{Fig3} illustrates the energy and angular momentum fluxes for EMRIs embedded in galaxies with a Dehnen type DM halo for different values of central density $\rho_s$ and scale radius $r_s$. As expected, both fluxes are negative, indicating that gravitational-wave emission causes the system to continuously lose energy and angular momentum. Based on the left panels of Fig.~\ref{Fig3}, the absolute value of the energy flux, $|dE/dt|$, decreases with increasing $\rho_s$ and $r_s$, representing that denser and more extended halos reduce the efficiency of gravitational-wave energy extraction, leading to a slower inspiral. In contrast, the absolute value of the angular momentum flux, $|dL/dt|$, shows a much weaker dependence on the halo parameters and exhibits only a slight enhancement for larger $\rho_s$ and $r_s$. This behavior implies that the effect of the DM halo is less important for the angular momentum flux as compared to the energy flux.  Furthermore, the influence of DM on energy flux is most pronounced at small $p/M$, whereas at larger separations the differences become negligible. 

\begin{figure*}[htp]
	\centering
	\includegraphics[width=0.45\textwidth]{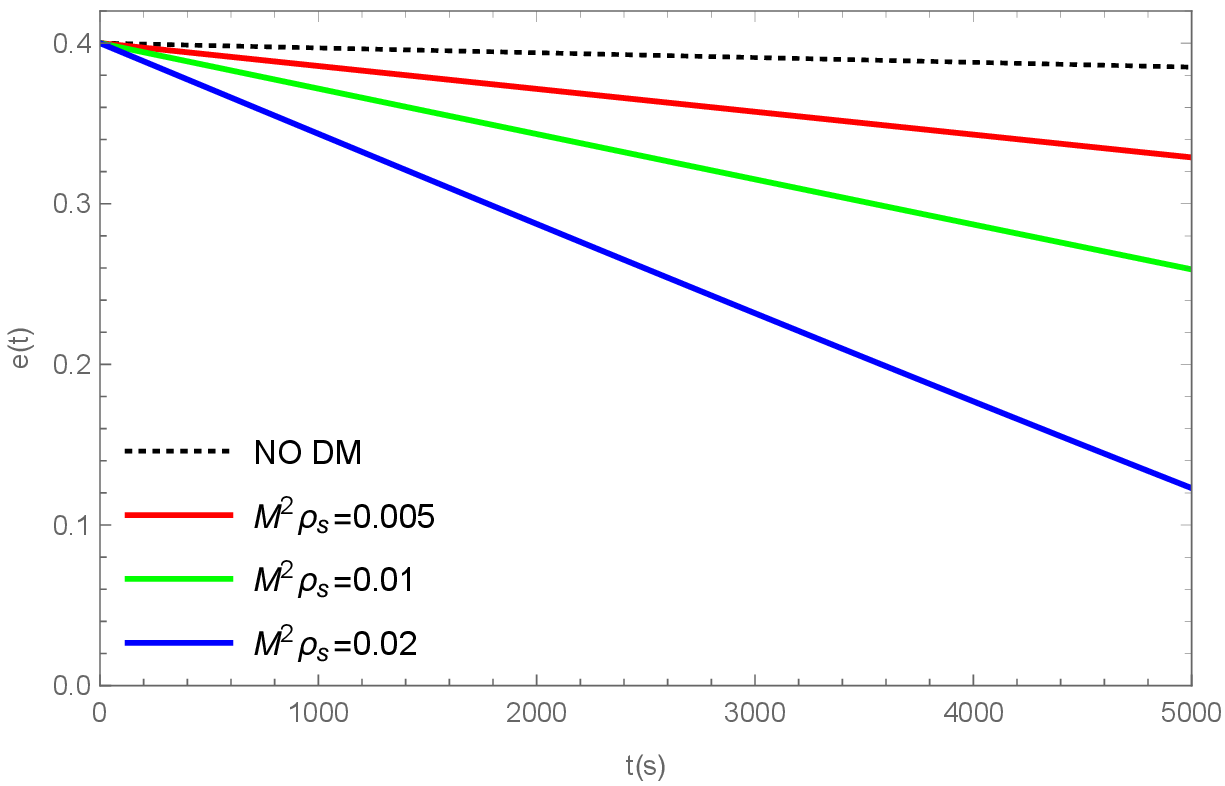}
	\includegraphics[width=0.45\textwidth]{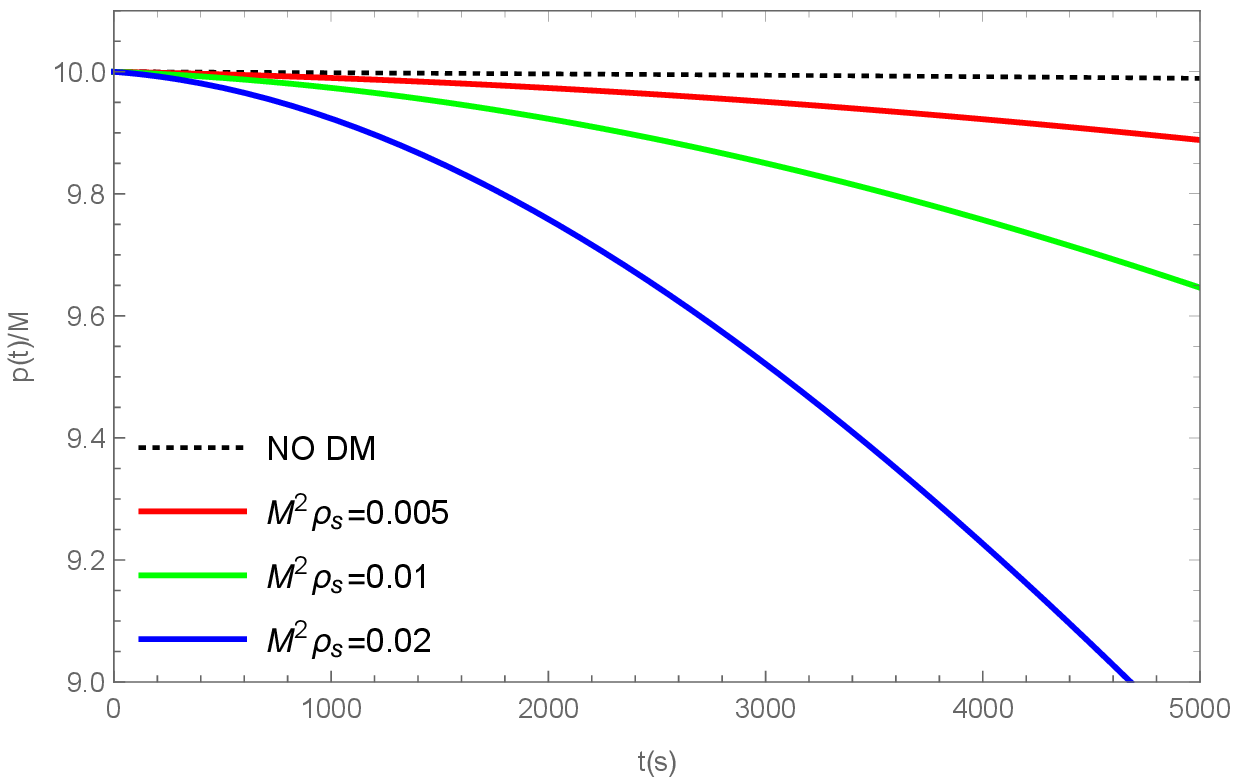}
	\includegraphics[width=0.45\textwidth]{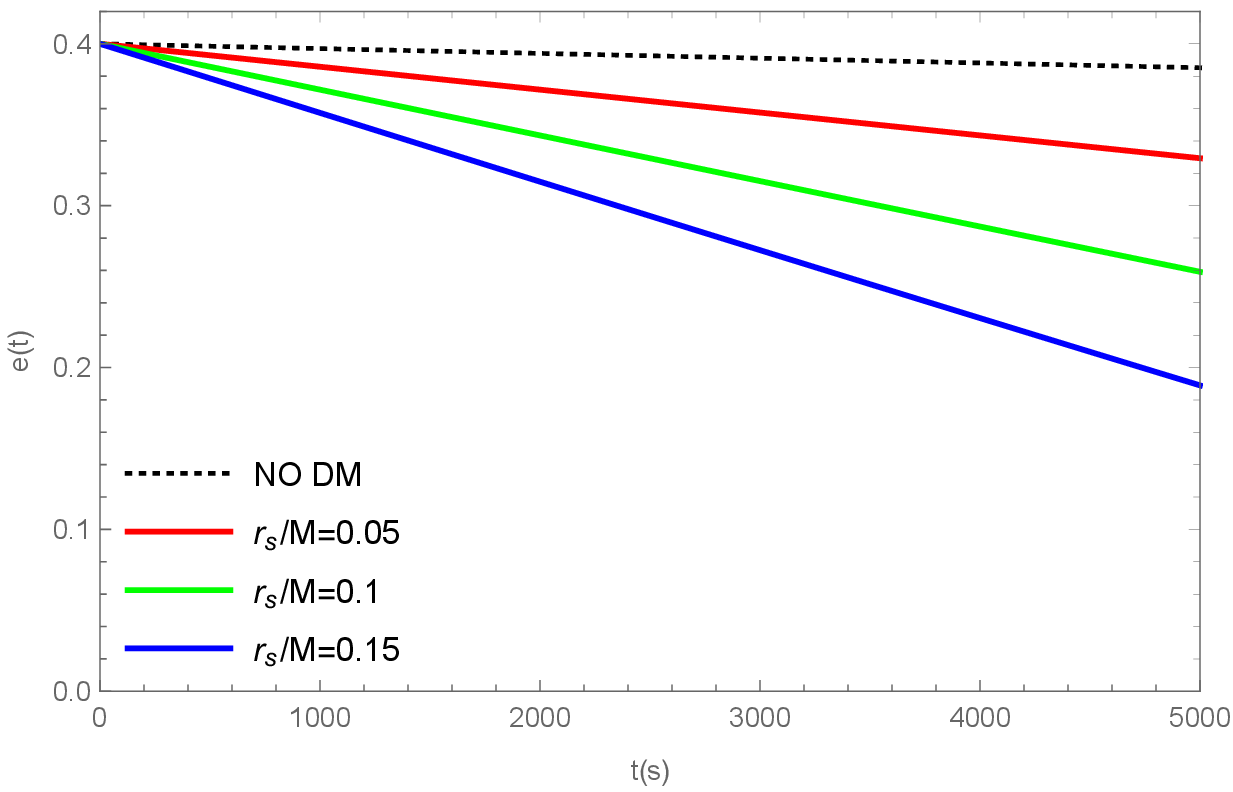}
	\includegraphics[width=0.45\textwidth]{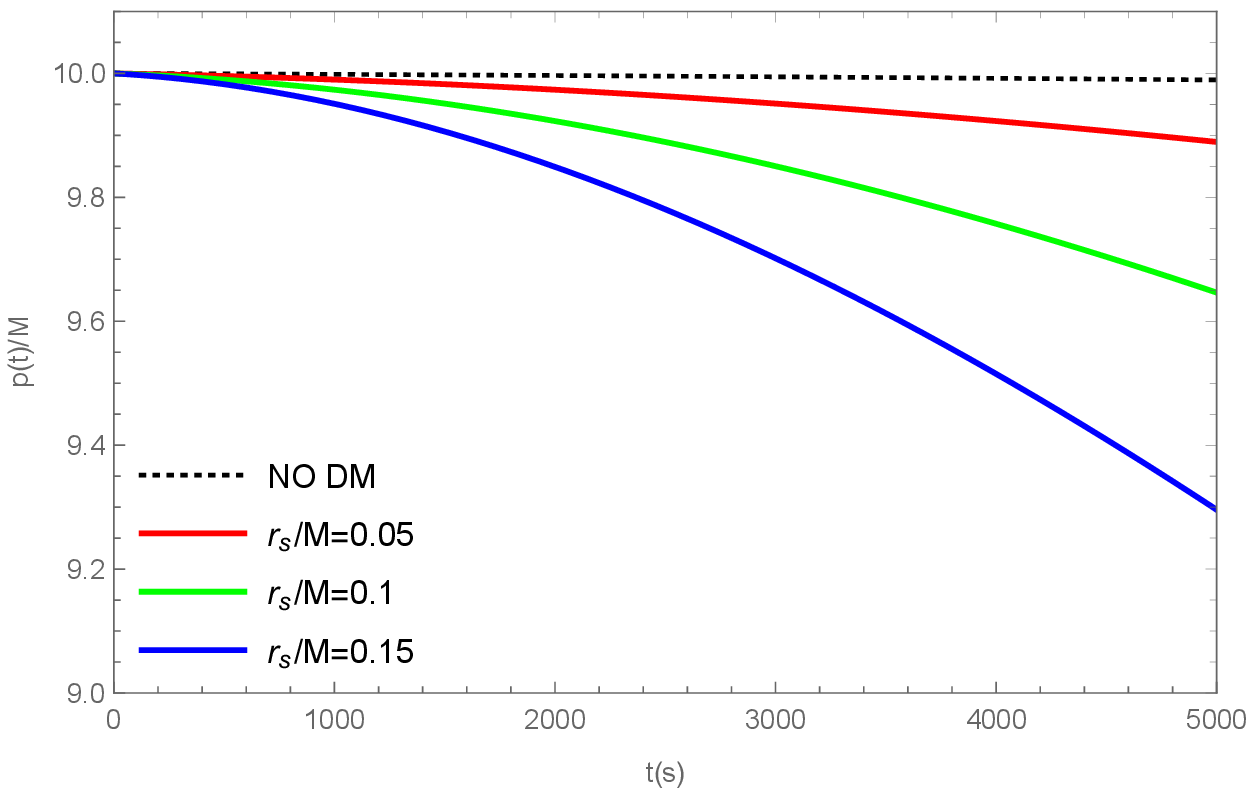}
	\caption{The evolution of the orbital parameters  $ e(t) $ and  $ p(t) $ for different values of $ \rho_{s} $ and $ r_{s}$.}
	\label{Fig4}
\end{figure*}

Using Eqs. \eqref{epsilon}-\eqref{ell} and \eqref{dEdtacc}-\eqref{dLdtorb}, the rate of change of the eccentricity and semi-latus rectum are given by \cite{Cao:2024a}
\begin{eqnarray}
	\label{eq:dedt}
	\frac{de}{dt} &=& \frac{e^2-1}{2e} \left( \frac{dE/dt}{E} + 2 \frac{dL/dt}{L}  \right),\\
	\label{eq:dpdt}
	\frac{dp}{dt} &=&\frac{da}{dt}\left(1-e^2 \right) -2ae\frac{de}{dt},
\end{eqnarray}
where $ \frac{da}{dt}=\frac{dE/dt}{dE/da} $. Fig.~\ref{Fig4} shows how eccentricity $e(t)$ and semi-latus rectum $p(t)$ evolve over time for EMRIs in a Dehnen type DM halo, as functions of the halo parameters $\rho_s$ and $r_s$. In particular, both $e(t)$ and $p(t)$ are shown to decrease monotonically with time, but the rate of decline is steeper for larger values of $\rho_s$ and $r_s$. This behavior reflects the increased gravitational potential due to the DM halo, accelerating the inspiral process. As a result, the orbit circularizes and shrinks more rapidly in halos with higher density normalization and scale radius, emphasizing the critical role of halo properties in shaping EMRI orbital evolution.


\section{Gravitational waveforms}\label{sezione4}

After determining the orbital evolution of the EMRI, the corresponding GW signal can be obtained within the framework of the leading-order quadrupole approximation. At this level, the strain tensor is expressed as
\begin{equation}
	h_{ij} = \frac{4 G \eta \mathcal{M}}{c^4 D_L} \left( v_i v_j - \frac{G \mathcal{M}}{r} n_i n_j \right),
\end{equation}
where $D_L$ denotes the luminosity distance between the source and the detector, $\eta = M \mu / \mathcal{M}^2$ represents the symmetric mass ratio, $\mathcal{M} = M + \mu$ the total mass of the system, $v_i$ the relative velocity components, and $n_i$ the components of the unit vector pointing along the separation of the compact objects. 

To extract the two independent polarization modes of the GW, we employ a detector-oriented orthonormal basis defined by
\begin{equation}
	\begin{aligned}
		\mathbf{e}_X &= (\cos \zeta, -\sin \zeta, 0), \\
		\mathbf{e}_Y &= (\cos \iota \sin \zeta, \cos \iota \cos \zeta, -\sin \iota), \\
		\mathbf{e}_Z &= (\sin \iota \sin \zeta, \sin \iota \cos \zeta, \cos \iota).
	\end{aligned}
\end{equation}
where $\iota$ denotes the inclination angle of the orbital plane with respect to the line of sight and $\zeta$ is the longitude of the pericenter. 

\begin{figure*}[htp]
	\centering
	\includegraphics[width=0.45\textwidth]{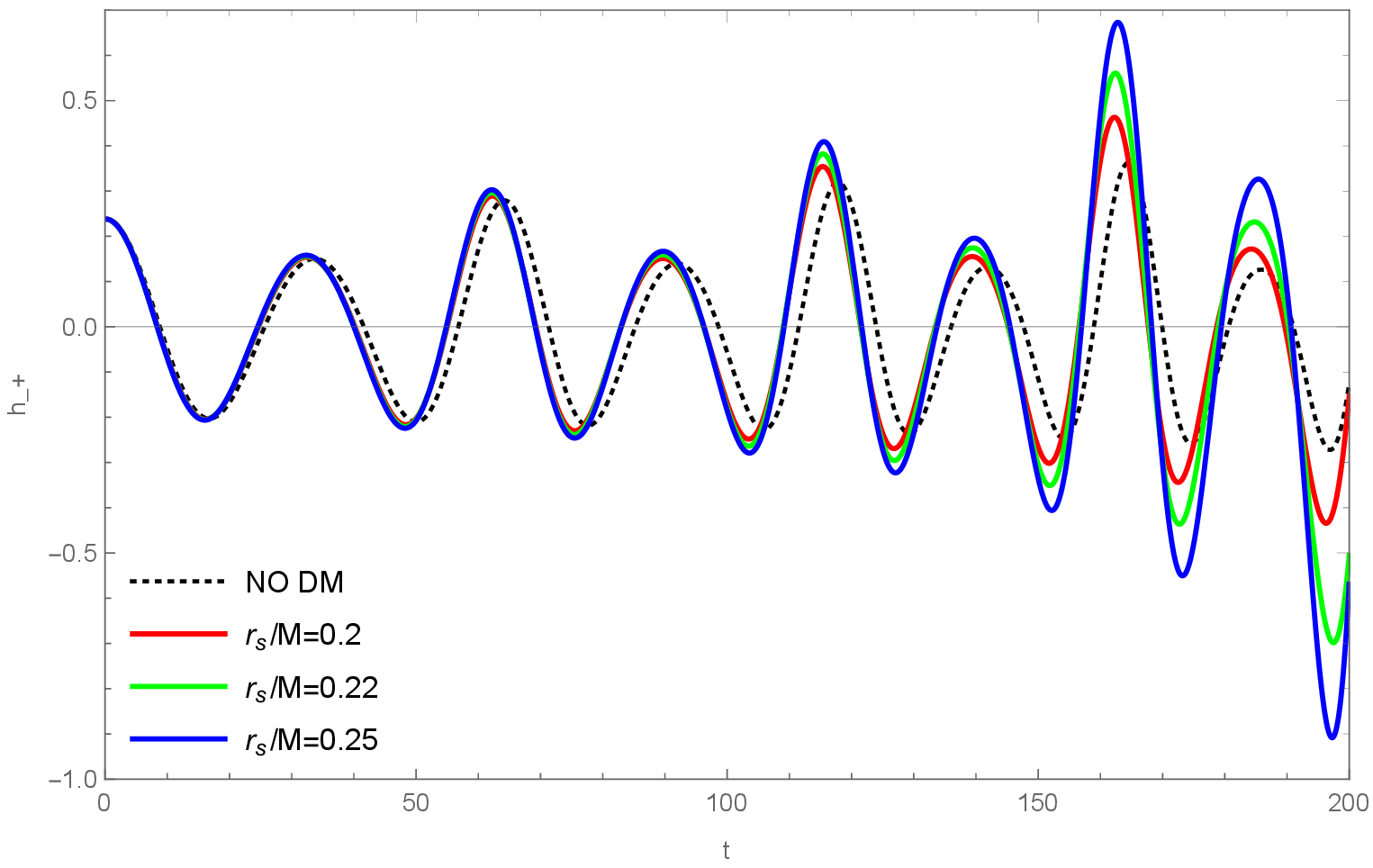}
	\includegraphics[width=0.45\textwidth]{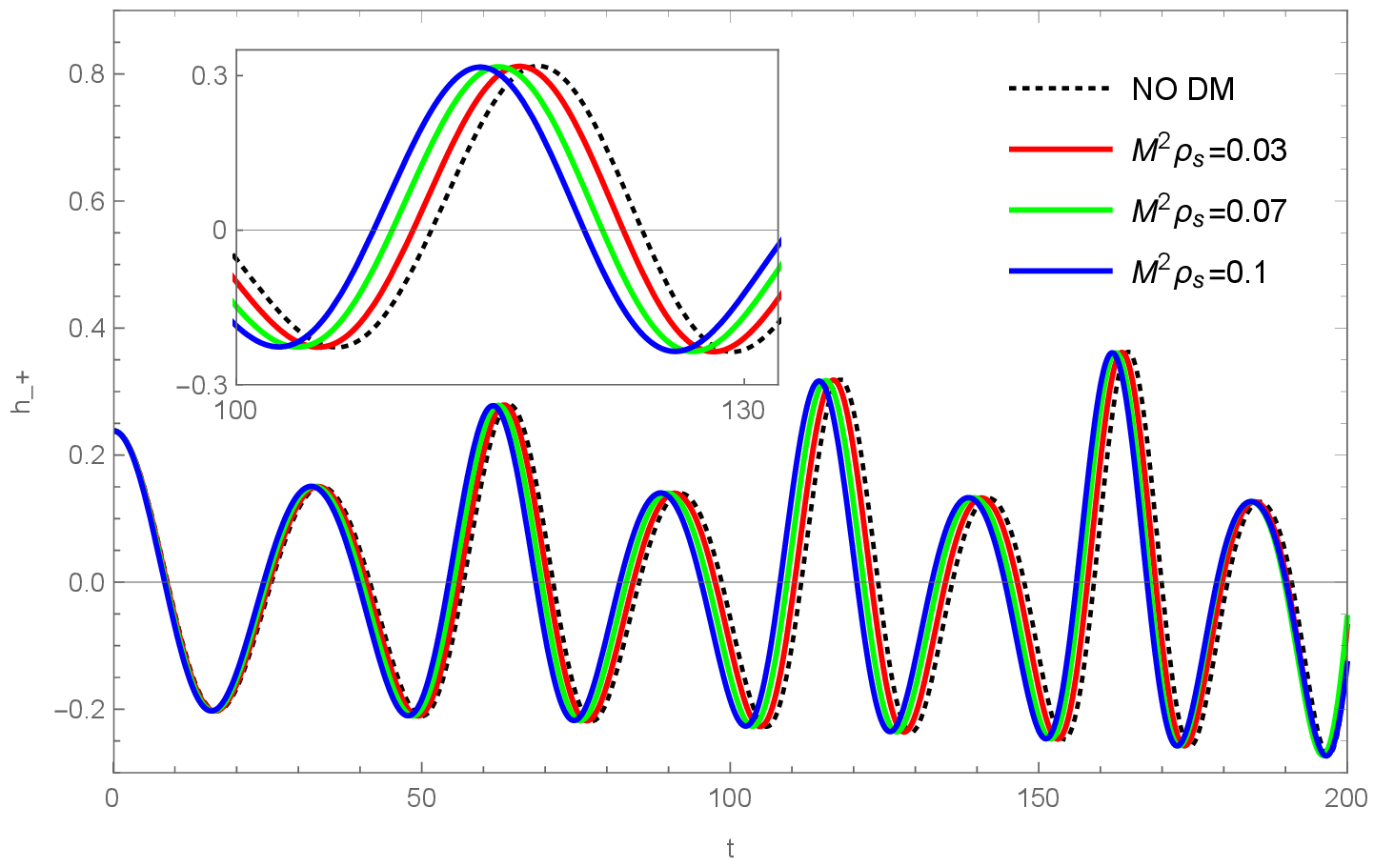}
	\caption{Time-domain $h_+$ waveforms for EMRIs embedded in a Dehnen type DM halo with varying $\rho_s$ and $r_s$.}
	\label{Fig5}
\end{figure*}

In the transverse-traceless gauge, the two polarizations of the GW take the form
\begin{equation}
	\begin{aligned}
		h_+ &= -\frac{2 \eta (G \mathcal{M})^2}{c^4 D_L r} (1 + \cos^2 \iota) \cos(2 \phi + 2 \zeta),\\
		h_\times &= -\frac{4 \eta (G \mathcal{M})^2}{c^4 D_L r} \cos \iota \, \sin(2 \phi + 2 \zeta),
	\end{aligned}
\end{equation}
where $r$ and $\phi$ denote the orbital separation and phase, respectively. The dependence of the $+$-mode gravitational waveforms on the DM parameters is demonstrated in Fig.~\ref{Fig5}. The right panel shows the effect of varying the halo density $\rho_{s}$ for a fixed $r_{s}$, revealing that as $\rho_s$ increases, the stronger gravitational potential and dynamical friction from the denser halo lead to a more pronounced dephasing compared to the vacuum case. This manifests as a phase acceleration, where oscillations arrive earlier as the DM density grows. The left panel shows the impact of changing the scale radius $r_{s}$ while keeping the halo normalization fixed, indicating that a smaller $r_{s}$ causes a smaller deviation of the vacuum waveform, while a larger $r_{s}$ has a greater effect on both the phase and amplitude of the waveform. 
Although the changes induced by DM are subtle during the early inspiral, they accumulate over time, leading to observable phase differences, which are critical to the accurate modeling of EMRI waveforms in DM.

\begin{figure*}[htp]
	\centering
	\includegraphics[width=0.45\textwidth]{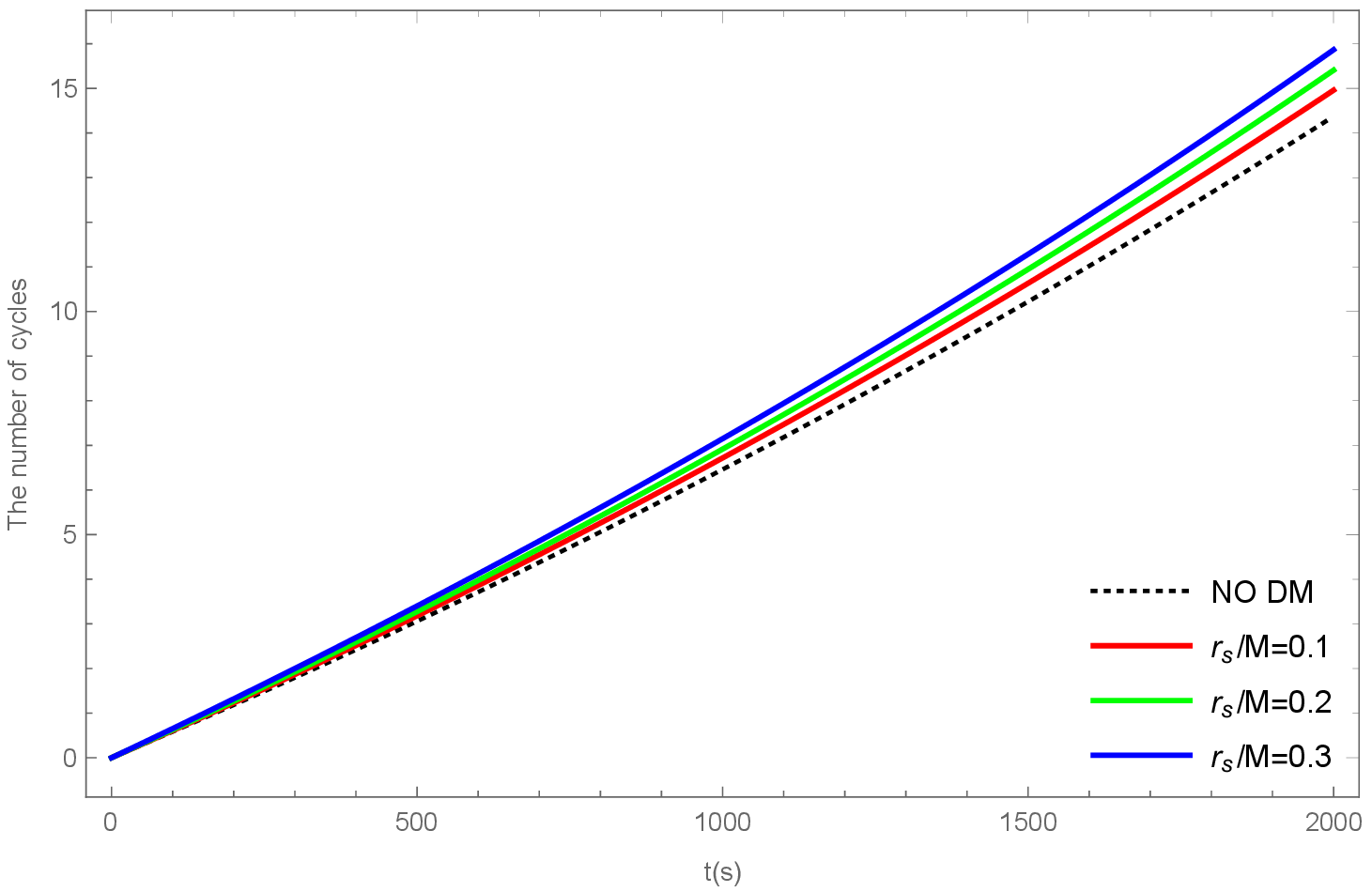}
	\includegraphics[width=0.45\textwidth]{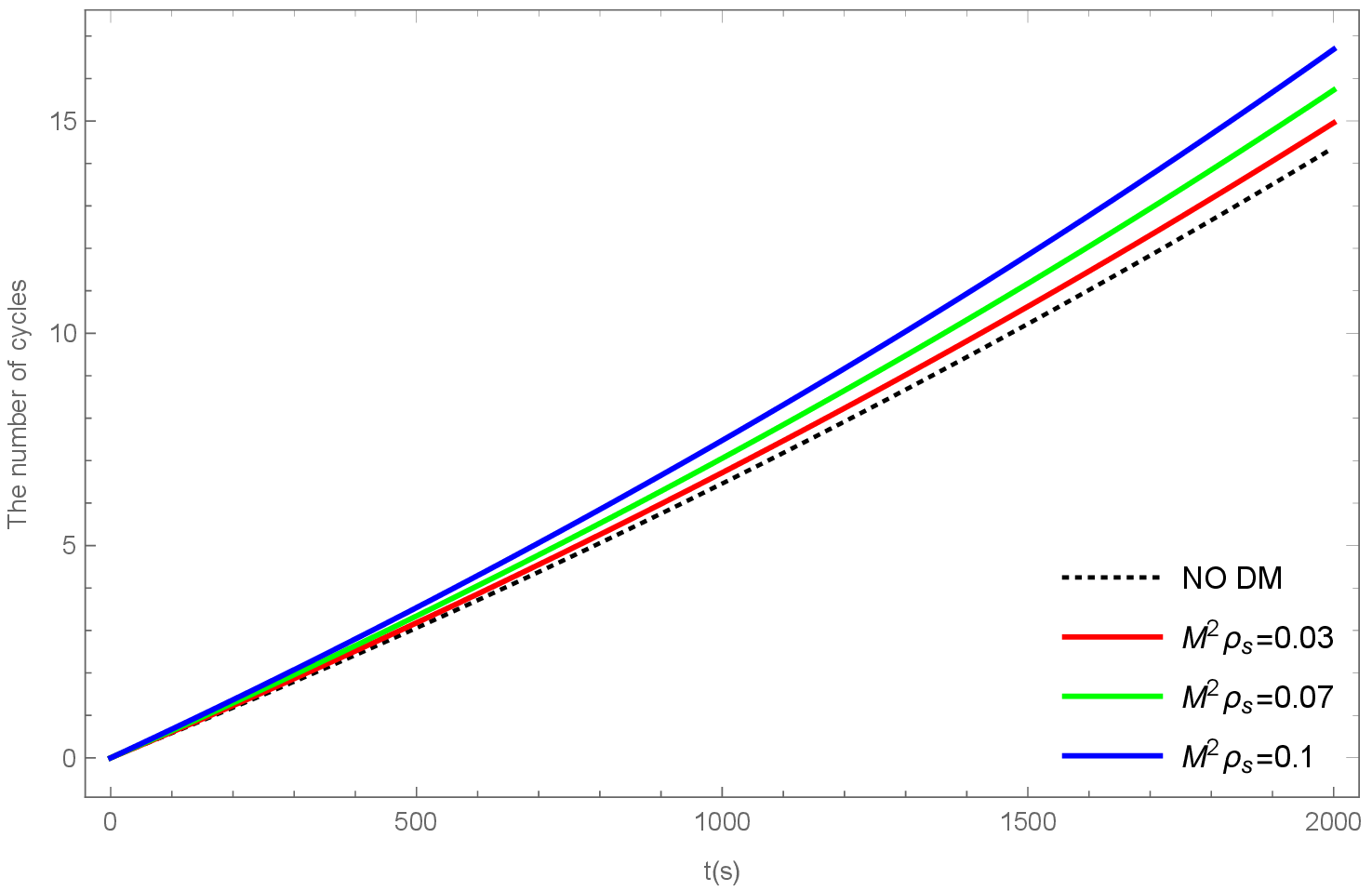}
	\caption{Evolution of the number of orbital cycles $\mathcal{N}(t)$ for EMRIs in the presence of a Dehnen DM halo, compared to the vacuum (No DM) case.}
	\label{Fig6}
\end{figure*}

To assess the effect of a Dehnen DM halo on the evolution of EMRIs, we compute the number of orbital cycles accumulated during the inspiral as \cite{Maselli:2020scalar,Barsanti:2022eccentric}
\begin{equation}
	\mathcal{N} =\frac{1}{\pi} \int_{t_1}^{t_2} \dot{\phi}(t) \, dt.
	\label{eq:Ncycles}
\end{equation}
Using Eqs.~(\ref{orbital3}) - (\ref{orbital2}), and (\ref{eq:Ncycles}), we evaluate \(\mathcal{N}(t)\) numerically for the EMRI model under consideration. Fig.~\ref{Fig6} shows the resulting \(\mathcal{N}(t)\) for EMRIs with and without the DM halo, indicating that the presence of DM produces a slight reduction in 
\(\mathcal{N}(t)\) relative to the vacuum case. This effect becomes more pronounced for halos that are more concentrated (smaller 
\(r_s\)) and denser (larger \(\rho_s\)), as the additional gravitational potential and dynamical effects from the halo slow down the system evolution. These findings suggest that dense or extended DM environments around EMRIs may alter their GW signature.

\section{Final remarks}\label{sezione5}

In this work, we have examined how the DM halo of the Dehnen type affects the dynamics and GW signatures of EMRIs. 
 After establishing the theoretical motivation for adopting the Dehnen profile as the dark-matter background and outlining its key physical properties, we find that the presence of dark matter induces systematic deviations from the vacuum case across all aspects of EMRI evolution, potentially yielding astrophysically significant effects over long inspiral timescales. We started with orbital period and pericenter precession, and we showed that both of these quantities are sensitive to the properties of the halo. Higher central density $\rho_s$ and greater radius of $r_s$ result in shorter orbital periods but more intense pericenteral precession $\Delta\phi$. Although variations in halo density $\rho_s$ cause slight changes, the scale radius has a stronger effect on orbital period at large distances and precession at small distances.

We then analyzed how radiation fluxes depend on DM parameters, and we noticed that as $\rho_s$ and $r_s$ increase, the absolute value of the energy flux $|dE/dt|$ decreases, which means that it becomes less efficient to extract energy from the orbit, thus slowing down the inspiral. On the other hand, the angular momentum flux $|dL/dt|$ is only slightly increased under the same conditions, indicating that DM suppresses energy dissipation more strongly than angular momentum loss.  Our analysis of orbital elements $e(t)$ and $p(t)$ demonstrated that the presence of DM speeds up circularization and contraction. Higher values of $\rho_s$ and $r_s$ result in a steeper drop in both parameters, underlining the role of halo properties in shaping the inspiratory path.

We continued our study by examining GW in the time domain and found that they exhibit slight but cumulative changes. While the initial inspiral signal remains close to the one in vacuum, the DM induces a systematic phase shift that becomes more pronounced as time progresses. Halos with larger $\rho_s$ and $r_s$ have stronger effects, leading to measurable phase shifts which may affect the estimation of parameters by detectors such as LISA.  Finally, we examined the number of orbital cycles and found that it is reduced in the presence of DM, particularly for denser and more concentrated halos. These results demonstrate that DM distributions around massive black holes can significantly influence EMRI evolution and GW signals; accurate modeling of these effects will be crucial for interpreting future space-based GW observations.

\acknowledgments A.A. thanks the University of Bologna theory group for their hospitality, and is particularly grateful to Silvia Pascoli for her support during his visit. The visit was funded by the European Union's Horizon Europe research and innovation programme under Marie Skłodowska
-Curie Staff Exchange grant agreement No 101086085-ASYMMETRY. He also acknowledges support from Mahmoud Safari and the Iranian Science Foundation, grant 4031449.
R.C.~is partially supported by the INFN grant FLAG, and his work has also been carried out in the framework of activities of the
National Group of Mathematical Physics (GNFM, INdAM) and the COST Action CA23115 (RQI). O.L.~acknowledges financial support from the Fondazione  ICSC, Spoke 3 Astrophysics and Cosmos Observations. National Recovery and Resilience Plan (Piano Nazionale di Ripresa e Resilienza, PNRR) Project ID CN$\_$00000013 "Italian Research Center on  High-Performance Computing, Big Data and Quantum Computing"  funded by MUR Missione 4 Componente 2 Investimento 1.4: Potenziamento strutture di ricerca e creazione di "campioni nazionali di R$\&$S (M4C2-19 )" - Next Generation EU (NGEU)
GRAB-IT Project, PNRR Cascade Funding
Call, Spoke 3, INAF Italian National Institute for Astrophysics, Project code CN00000013, Project Code (CUP): C53C22000350006, cost center STI442016.

\bibliographystyle{JHEP}

\bibliography{mybib.bib}

@article{Navarro:1995dmo,
    author = "Navarro, Julio F. and Frenk, Carlos S. and White, Simon D. M.",
    title = "{The Structure of Cold Dark Matter Halos}",
    journal = "Astrophys. J.",
    volume = "462",
    pages = "563",
    year = "1996",
    doi = "10.1086/177173",
    eprint = "astro-ph/9508025",
    archivePrefix = "arXiv"
}

@article{Dehnen:1993,
    author = "Dehnen, Walter",
    title = "{A Family of Potential-Density Pairs for Spherical Galaxies and Bulges}",
    journal = "Mon. Not. R. Astron. Soc.",
    volume = "265",
    pages = "250",
    year = "1993",
    doi = "10.1093/mnras/265.1.250"
}

@article{Stegmann:2020,
    author = "Stegmann, J. and Capelo, P. R. and Bortolas, E. and Mayer, L.",
    title = "{Improved constraints from ultra-faint dwarf galaxies on primordial black holes as dark matter}",
    journal = "Mon. Not. R. Astron. Soc.",
    volume = "492",
    pages = "5247",
    year = "2020",
    doi = "10.1093/mnras/staa170",
    eprint = "1910.04793",
    archivePrefix = "arXiv",
    primaryClass = "astro-ph.GA"
}

@article{Hamil:2024dehnen,
    author = "Hamil, B. and Al-Badawi, Ahmad and Lütfüoğlu, B. C.",
    title = "{Geodesics and scalar perturbations of Schwarzschild black holes embedded in a Dehnen-type dark matter halo with quintessence}",
    eprint = "2505.18611",
    archivePrefix = "arXiv",
    primaryClass = "gr-qc",
    year = "2025",
    journal = "arXiv preprint",
}

@article{Pantig:2022dehnen,
    author = "Pantig, Reggie C. and Övgün, Ali",
    title = "{Dehnen halo effect on a black hole in an ultra-faint dwarf galaxy}",
    journal = "JCAP",
    volume = "08",
    number = "08",
    pages = "056",
    year = "2022",
    doi = "10.1088/1475-7516/2022/08/056",
    eprint = "2202.07404",
    archivePrefix = "arXiv",
    primaryClass = "astro-ph.GA"
}

@article{Gohain:2024dark,
    author = "Gohain, Mrinnoy M. and Phukon,  Prabwal and Bhuyan, Kalyan",
    title = "{Thermodynamics and Null Geodesics of a Schwarzschild Black Hole Surrounded by a Dehnen Type Dark Matter Halo}",
    journal = "Phys. Dark Univ.",
    volume = "46",
    pages = "101683",
    year = "2024",
    doi = "10.1016/j.dark.2024.101683",
    eprint = "2407.02872",
    archivePrefix = "arXiv",
    primaryClass = "gr-qc"
}

@article{Ali:2025dehnen,
    author = "Ali, Ahmed and Molla, Niyaz U. and Ghosh, Sumanta G. and Ramasamy, A. and Debnath, Ujjal",
    title = "{Gravitational lensing and dark matter profiles: Testing Dehnen models with black hole shadows}",
    journal = "Phys. Dark Univ.",
    volume = "48",
    pages = "101859",
    year = "2025",
    doi = "10.1016/j.dark.2025.101859"
}

@article{LIGOScientific:2016aoc,
    author = "Abbott, B. P. and others",
    collaboration = "LIGO Scientific, Virgo",
    title = "{Observation of Gravitational Waves from a Binary Black Hole Merger}",
    eprint = "1602.03837",
    archivePrefix = "arXiv",
    primaryClass = "gr-qc",
    reportNumber = "LIGO-P150914",
    doi = "10.1103/PhysRevLett.116.061102",
    journal = "Phys. Rev. Lett.",
    volume = "116",
    number = "6",
    pages = "061102",
    year = "2016"
}

@article{LISA:2022kgy,
    author = "Arun, K. G. and others",
    collaboration = "LISA",
    title = "{New horizons for fundamental physics with LISA}",
    eprint = "2205.01597",
    archivePrefix = "arXiv",
    primaryClass = "gr-qc",
    doi = "10.1007/s41114-022-00036-9",
    journal = "Living Rev. Rel.",
    volume = "25",
    number = "1",
    pages = "4",
    year = "2022"
}

@article{Dehnen:1993uh,
    author = "Dehnen, W.",
    title = "{A Family of Potential-Density Pairs for Spherical Galaxies and Bulges}",
    journal = "Mon. Not. Roy. Astron. Soc.",
    volume = "265",
    pages = "250",
    year = "1993"
}

@article{Stegmann:2019wyz,
    author = "Stegmann, Jakob and Capelo, Pedro R. and Bortolas, Elisa and Mayer, Lucio",
    title = "{Improved constraints from ultra-faint dwarf galaxies on primordial black holes as dark matter}",
    eprint = "1910.04793",
    archivePrefix = "arXiv",
    primaryClass = "astro-ph.GA",
    doi = "10.1093/mnras/staa170",
    journal = "Mon. Not. Roy. Astron. Soc.",
    volume = "492",
    number = "4",
    pages = "5247--5260",
    year = "2020"
}

@article{Inoue:2017voc,
    author = "Inoue, Shigeki",
    title = "{Emergence of a stellar cusp by a dark matter cusp in a low-mass compact ultrafaint dwarf galaxy}",
    eprint = "1701.07459",
    archivePrefix = "arXiv",
    primaryClass = "astro-ph.GA",
    doi = "10.1093/mnras/stx393",
    journal = "Mon. Not. Roy. Astron. Soc.",
    volume = "467",
    number = "4",
    pages = "4491--4500",
    year = "2017"
}

@article{errani2018systematics,
    author = "Errani, Rapha{\"e}l and Pe{\~n}arrubia, Jorge and Walker, Matthew G",
    title = "{Systematics in virial mass estimators for pressure-supported systems}",
    eprint = "1805.00484",
    archivePrefix = "arXiv",
    primaryClass = "astro-ph",
    doi = "10.1093/mnras/sty2505",
    journal = "Mon. Not. Roy. Astron. Soc.",
    volume = "481",
    number = "4",
    pages = "5073--5090",
    year = "2018"
}

@article{Bustamante-Rosell:2021ldj,
    author = "Bustamante-Rosell, Maria Jose and Noyola, Eva and Gebhardt, Karl and Fabricius, Maximilian H. and Mazzalay, Ximena and Thomas, Jens and Zeimann, Greg",
    title = "{Dynamical Analysis of the Dark Matter and Central Black Hole Mass in the Dwarf Spheroidal Leo I}",
    eprint = "2111.04770",
    archivePrefix = "arXiv",
    primaryClass = "astro-ph.GA",
    doi = "10.3847/1538-4357/ac0c79",
    journal = "Astrophys. J.",
    volume = "921",
    number = "2",
    pages = "107",
    year = "2021"
}

@article{Pantig:2022whj,
    author = {Pantig, Reggie C. and \"Ovg\"un, Ali},
    title = "{Dehnen halo effect on a black hole in an ultra-faint dwarf galaxy}",
    eprint = "2202.07404",
    archivePrefix = "arXiv",
    primaryClass = "astro-ph.GA",
    doi = "10.1088/1475-7516/2022/08/056",
    journal = "JCAP",
    volume = "08",
    number = "08",
    pages = "056",
    year = "2022"
}

@article{Al-Badawi:2024asn,
    author = "Al-Badawi, Ahmad and Shaymatov, Sanjar and Sekhmani, Yassine",
    title = "{Schwarzschild black hole in galaxies surrounded by a dark matter halo}",
    eprint = "2411.01145",
    archivePrefix = "arXiv",
    primaryClass = "gr-qc",
    doi = "10.1088/1475-7516/2025/02/014",
    journal = "JCAP",
    volume = "02",
    pages = "014",
    year = "2025"
}

@article{Gohain:2024eer,
    author = "Gohain, Mrinnoy M. and Phukon, Prabwal and Bhuyan, Kalyan",
    title = "{Thermodynamics and null geodesics of a Schwarzschild black hole surrounded by a Dehnen type dark matter halo}",
    eprint = "2407.02872",
    archivePrefix = "arXiv",
    primaryClass = "gr-qc",
    doi = "10.1016/j.dark.2024.101683",
    journal = "Phys. Dark Univ.",
    volume = "46",
    pages = "101683",
    year = "2024"
}

@article{Al-Badawi:2024qpt,
    author = "Al-Badawi, Ahmad and Shaymatov, Sanjar",
    title = "{Quasinormal modes and shadow of Schwarzschild black holes embedded in a Dehnen-type dark matter halo exhibiting a cloud of strings}",
    eprint = "2412.20037",
    archivePrefix = "arXiv",
    primaryClass = "gr-qc",
    doi = "10.1088/1572-9494/ad89b2",
    journal = "Commun. Theor. Phys.",
    volume = "77",
    number = "3",
    pages = "035402",
    year = "2025"
}

@article{Al-Badawi:2025njy,
    author = "Al-Badawi, Ahmad and Shaymatov, Sanjar",
    title = "{Astrophysical properties of static black holes embedded in a Dehnen type dark matter halo with the presence of quintessential field*}",
    eprint = "2501.15397",
    archivePrefix = "arXiv",
    primaryClass = "gr-qc",
    doi = "10.1088/1674-1137/adb2fd",
    journal = "Chin. Phys. C",
    volume = "49",
    number = "5",
    pages = "055101",
    year = "2025"
}

@article{Hughes:2024tja,
    author = "Hughes, Scott A.",
    title = "{Parameterizing black hole orbits for adiabatic inspiral}",
    eprint = "2401.09577",
    archivePrefix = "arXiv",
    primaryClass = "gr-qc",
    doi = "10.1103/PhysRevD.109.064077",
    journal = "Phys. Rev. D",
    volume = "109",
    number = "6",
    pages = "064077",
    year = "2024"
}

@article{Barack:2022pde,
    author = "Barack, Leor and Long, Oliver",
    title = "{Self-force correction to the deflection angle in black-hole scattering: A scalar charge toy model}",
    eprint = "2209.03740",
    archivePrefix = "arXiv",
    primaryClass = "gr-qc",
    doi = "10.1103/PhysRevD.106.104031",
    journal = "Phys. Rev. D",
    volume = "106",
    number = "10",
    pages = "104031",
    year = "2022"
}

@article{Cardoso:2020iji,
    author = "Cardoso, Vitor and Macedo, Caio F. B. and Vicente, Rodrigo",
    title = "{Eccentricity evolution of compact binaries and applications to gravitational-wave physics}",
    eprint = "2010.15151",
    archivePrefix = "arXiv",
    primaryClass = "gr-qc",
    doi = "10.1103/PhysRevD.103.023015",
    journal = "Phys. Rev. D",
    volume = "103",
    number = "2",
    pages = "023015",
    year = "2021"
}

@article{Cardoso:2021wlq,
    author = "Cardoso, Vitor and Destounis, Kyriakos and Duque, Francisco and Macedo, Rodrigo Panosso and Maselli, Andrea",
    title = "{Black holes in galaxies: Environmental impact on gravitational-wave generation and propagation}",
    eprint = "2109.00005",
    archivePrefix = "arXiv",
    primaryClass = "gr-qc",
    doi = "10.1103/PhysRevD.105.L061501",
    journal = "Phys. Rev. D",
    volume = "105",
    number = "6",
    pages = "L061501",
    year = "2022"
}

@article{Cardoso:2022whc,
    author = "Cardoso, Vitor and Destounis, Kyriakos and Duque, Francisco and Panosso Macedo, Rodrigo and Maselli, Andrea",
    title = "{Gravitational Waves from Extreme-Mass-Ratio Systems in Astrophysical Environments}",
    eprint = "2210.01133",
    archivePrefix = "arXiv",
    primaryClass = "gr-qc",
    doi = "10.1103/PhysRevLett.129.241103",
    journal = "Phys. Rev. Lett.",
    volume = "129",
    number = "24",
    pages = "241103",
    year = "2022"
}

@article{Figueiredo:2023gas,
    author = "Figueiredo, Enzo and Maselli, Andrea and Cardoso, Vitor",
    title = "{Black holes surrounded by generic dark matter profiles: Appearance and gravitational-wave emission}",
    eprint = "2303.08183",
    archivePrefix = "arXiv",
    primaryClass = "gr-qc",
    doi = "10.1103/PhysRevD.107.104033",
    journal = "Phys. Rev. D",
    volume = "107",
    number = "10",
    pages = "104033",
    year = "2023"
}

@article{Dai:2021olt,
    author = "Dai, Ning and Gong, Yungui and Jiang, Tong and Liang, Dicong",
    title = "{Intermediate mass-ratio inspirals with dark matter minispikes}",
    eprint = "2111.13514",
    archivePrefix = "arXiv",
    primaryClass = "gr-qc",
    doi = "10.1103/PhysRevD.106.064003",
    journal = "Phys. Rev. D",
    volume = "106",
    number = "6",
    pages = "064003",
    year = "2022"
}

@article{Luongo:2023aib,
    author = "Luongo, Orlando and Quevedo, Hernando",
    title = "{Repulsive gravity in regular black holes}",
    eprint = "2305.11185",
    archivePrefix = "arXiv",
    primaryClass = "gr-qc",
    doi = "10.1088/1361-6382/ad4ae4",
    journal = "Class. Quant. Grav.",
    volume = "41",
    number = "12",
    pages = "125011",
    year = "2024"
}

@article{Luongo:2014qoa,
    author = "Luongo, Orlando and Quevedo, Hernando",
    title = "{Characterizing repulsive gravity with curvature eigenvalues}",
    eprint = "1407.1530",
    archivePrefix = "arXiv",
    primaryClass = "gr-qc",
    doi = "10.1103/PhysRevD.90.084032",
    journal = "Phys. Rev. D",
    volume = "90",
    number = "8",
    pages = "084032",
    year = "2014"
}

@article{Chu:2021uec,
    author = "Chu, Chong-Sun and Tan, Hai Siong",
    title = "{Generalized Darmois{\textendash}Israel Junction Conditions}",
    eprint = "2103.06314",
    archivePrefix = "arXiv",
    primaryClass = "hep-th",
    doi = "10.3390/universe8050250",
    journal = "Universe",
    volume = "8",
    number = "5",
    pages = "250",
    year = "2022"
}

@article{Boshkayev:2020kle,
    author = "Boshkayev, Kuantay and Idrissov, Anuar and Luongo, Orlando and Malafarina, Daniele",
    title = "{Accretion disc luminosity for black holes surrounded by dark matter}",
    eprint = "2006.01269",
    archivePrefix = "arXiv",
    primaryClass = "astro-ph.HE",
    doi = "10.1093/mnras/staa1564",
    journal = "Mon. Not. Roy. Astron. Soc.",
    volume = "496",
    number = "2",
    pages = "1115--1123",
    year = "2020"
}

@article{Boshkayev:2021chc,
    author = "Boshkayev, Kuantay and Konysbayev, Talgar and Kurmanov, Ergali and Luongo, Orlando and Malafarina, Daniele and Quevedo, Hernando",
    title = "{Luminosity of accretion disks in compact objects with a quadrupole}",
    eprint = "2106.04932",
    archivePrefix = "arXiv",
    primaryClass = "gr-qc",
    doi = "10.1103/PhysRevD.104.084009",
    journal = "Phys. Rev. D",
    volume = "104",
    number = "8",
    pages = "084009",
    year = "2021"
}

@article{Capozziello:2025ycu,
    author = "Capozziello, Salvatore and Gambino, Serena and Luongo, Orlando",
    title = "{Comparing Bondi and Novikov{\textendash}Thorne accretion disk luminosity around regular black holes}",
    eprint = "2503.21987",
    archivePrefix = "arXiv",
    primaryClass = "gr-qc",
    doi = "10.1016/j.dark.2025.101950",
    journal = "Phys. Dark Univ.",
    volume = "48",
    pages = "101950",
    year = "2025"
}

@article{Dunsby:2016lkw,
    author = "Dunsby, Peter K. S. and Luongo, Orlando and Reverberi, Lorenzo",
    title = "{Dark Energy and Dark Matter from an additional adiabatic fluid}",
    eprint = "1604.06908",
    archivePrefix = "arXiv",
    primaryClass = "gr-qc",
    doi = "10.1103/PhysRevD.94.083525",
    journal = "Phys. Rev. D",
    volume = "94",
    number = "8",
    pages = "083525",
    year = "2016"
}

@article{Zhang:2024ugv,
    author = "Zhang, Chao and Fu, Guoyang and Dai, Ning",
    title = "{Detecting dark matter halos with extreme mass-ratio inspirals}",
    eprint = "2401.04467",
    archivePrefix = "arXiv",
    primaryClass = "gr-qc",
    doi = "10.1088/1475-7516/2024/04/088",
    journal = "JCAP",
    volume = "04",
    pages = "088",
    year = "2024"
}

@article{Shakeshaft:1974iau,
    author = "Shakeshaft, J. R. and others",
    title = "{The Formation and Dynamics of Galaxies}",
    doi = "https://link.springer.com/book/10.1007/978-94-010-2222-4",
    journal = "IAU Symposium, Vol.",
    volume = "58",
    year = "1974"}

@article{Macedo:2013a,
    author = "Macedo, C. F. B. and Pani, P. and Cardoso, V. and Crispino,  L. C. B.",
    title = "{Into the lair: gravitational-wave
signatures of dark matter}",
     eprint = "1302.2646",
    archivePrefix = "arXiv",
    primaryClass = "gr-qc",
    doi = "10.1088/0004-637X/774/1/48",
    journal = "Astrophys. J.",
    volume = "744",
    pages = "48",
    year = "2013"}

@article{Gondolo:1999ef,
    author = "Gondolo, Paolo and Silk, Joseph",
    title = "{Dark matter annihilation at the galactic center}",
    eprint = "astro-ph/9906391",
    archivePrefix = "arXiv",
    reportNumber = "MPI-PHT-99-10, OUAST-99-9",
    doi = "10.1103/PhysRevLett.83.1719",
    journal = "Phys. Rev. Lett.",
    volume = "83",
    pages = "1719--1722",
    year = "1999"
}

@article{Chandrasekhar:1943,
       author ="Chandrasekhar, S.",
        title = "{Dynamical Friction. I. General Considerations: the Coefficient of Dynamical Friction}",
      journal = {Astrophys. J.},
         year = "1943",
        month = mar,
       volume = "97",
        pages = "255",
          doi = "10.1086/144517"}

@article{Eda:91,
    author = "Eda, K. and Itoh, Y. and Kuroyanagi, S. and Silk, J.",
    title = "{Gravitational waves as a probe of dark matter
minispikes}",
    eprint = "1408.3534",
    archivePrefix = "arXiv",
    primaryClass = "gr-qc",
    doi = "10.1103/PhysRevD.91.044045",
    journal = "Phys. Rev. D",
    volume = "91",
    pages = "044045",
    year = "2015"}

@article{Peters:131,
    author = "Peters, P. C. and Mathews, J.",
    title = "{Gravitational radiation
from point masses in a keplerian orbit}",
    doi = "10.1103/PhysRev.131.435",
    journal = "Phys. Rev.",
    volume = "131",
    pages = "435",
    year = "1963"}

@article{Cao:2024a,
    author = "Cao, Y. and Cheng, Y.-Z. and Li, G.-L. and Tang, Y.",
    title = "{Probing vector
gravitational atom with eccentric intermediate mass-ratio inspirals}",
     eprint = "2411.17247",
    archivePrefix = "arXiv",
    primaryClass = "gr-qc",
    doi = "10.1103/PhysRevD.111.083011",
    journal = "Phys. Rev. D",
    volume = "111",
    pages = "083011",
    year = "2025"}

@article{Xu:2018wow,
    author = "Xu, Zhaoyi and Hou, Xian and Gong, Xiaobo and Wang, Jiancheng",
    title = "{Black Hole Space-time In Dark Matter Halo}",
    eprint = "1803.00767",
    archivePrefix = "arXiv",
    primaryClass = "gr-qc",
    doi = "10.1088/1475-7516/2018/09/038",
    journal = "JCAP",
    volume = "09",
    pages = "038",
    year = "2018"
}

@article{Maselli:2020scalar,
    author = "Maselli, Andrea and Franchini, Nicola and Gualtieri, Leonardo and Sotiriou, Thomas P.",
    title = "{Detecting scalar fields with Extreme Mass Ratio Inspirals}",
    journal = "Phys. Rev. Lett.",
    volume = "125",
    number = "14",
    pages = "141101",
    year = "2020",
    doi = "10.1103/PhysRevLett.125.141101",
    eprint = "2004.11895",
    archivePrefix = "arXiv",
    primaryClass = "gr-qc"
}

@article{Barsanti:2022eccentric,
    author = "Barsanti, S. and Franchini, N. and Gualtieri, L. and Maselli, A. and Sotiriou, T. P.",
    title = "{Extreme mass-ratio inspirals as probes of scalar fields: Eccentric equatorial orbits around Kerr black holes}",
    journal = "Phys. Rev. D",
    volume = "106",
    number = "4",
    pages = "044029",
    year = "2022",
    doi = "10.1103/PhysRevD.106.044029",
    eprint = "2203.05003",
    archivePrefix = "arXiv",
    primaryClass = "gr-qc"
}

@article{Reines:2022ste,
    author = "Reines, Amy E.",
    title = "{Hunting for massive black holes in dwarf galaxies}",
    eprint = "2201.10569",
    archivePrefix = "arXiv",
    primaryClass = "astro-ph.GA",
    doi = "10.1038/s41550-021-01556-0",
    journal = "Nature Astron.",
    volume = "6",
    number = "1",
    pages = "26--34",
    year = "2022"
}

@article{Cadoni:2017evg,
    author = {Cadoni, Mariano and Casadio, Roberto and Giusti, Andrea and M{\"u}ck, Wolfgang and Tuveri, Matteo},
    title = "{Effective Fluid Description of the Dark Universe}",
    eprint = "1707.09945",
    archivePrefix = "arXiv",
    primaryClass = "gr-qc",
    doi = "10.1016/j.physletb.2017.11.058",
    journal = "Phys. Lett. B",
    volume = "776",
    pages = "242--248",
    year = "2018"
}

@article{Cadoni:2018dnd,
    author = "Cadoni, M. and Casadio, R. and Giusti, A. and Tuveri, M.",
    title = "{Emergence of a Dark Force in Corpuscular Gravity}",
    eprint = "1801.10374",
    archivePrefix = "arXiv",
    primaryClass = "gr-qc",
    doi = "10.1103/PhysRevD.97.044047",
    journal = "Phys. Rev. D",
    volume = "97",
    number = "4",
    pages = "044047",
    year = "2018"
}

@article{Giusti:2021shf,
    author = "Giusti, Andrea and Buffa, Silvia and Heisenberg, Lavinia and Casadio, Roberto",
    title = "{A quantum state for the late Universe}",
    eprint = "2108.05111",
    archivePrefix = "arXiv",
    primaryClass = "gr-qc",
    doi = "10.1016/j.physletb.2022.136900",
    journal = "Phys. Lett. B",
    volume = "826",
    pages = "136900",
    year = "2022"
}
\end{document}